\documentclass{llncs}

\usepackage{comment}
\usepackage{url}
\usepackage{etoolbox}
\usepackage{rotating}
\usepackage{balance}
\usepackage{enumerate}
\usepackage{paralist}
\usepackage{multirow}
\usepackage{listings}
\usepackage{color}
\usepackage{courier}
\usepackage[utf8]{inputenc}
\usepackage[colorinlistoftodos,textsize=scriptsize]{todonotes}
\usepackage[version=3]{mhchem} 
\usepackage[numbers,sort]{natbib}
\usepackage{subfig}
\usepackage{float}
\usepackage{amsmath}
\usepackage{amssymb}

\usepackage{aliascnt}
\newaliascnt{eqfloat}{equation}
\newfloat{eqfloat}{h}{eqflts}
\floatname{eqfloat}{Equation}

\newcommand*{\ORGeqfloat}{}
\let\ORGeqfloat\eqfloat
\def\eqfloat{%
  \let\ORIGINALcaption\caption
  \def\caption{%
    \addtocounter{equation}{-1}%
    \ORIGINALcaption
  }%
  \ORGeqfloat
}

\begin{document}
%
\newtoggle{epspaper}
\togglefalse{epspaper}

\title{Efficient High-Speed WPA2 Brute Force Attacks using Scalable Low-Cost FPGA Clustering}
\subtitle{Extended Version}

\author{Markus Kammerstetter\inst{1} \and Markus Muellner\inst{1} \and
Daniel Burian\inst{1} \and Christian Kudera\inst{1} \and Wolfgang Kastner\inst{2}}

\institute{Secure Systems Lab Vienna, Automation Systems Group, Institute of Computer Aided Automation, Vienna University of Technology,\\
\email{\{mk,mmuellner,dburian,ckudera\} @ seclab.tuwien.ac.at},\\
\and
Automation Systems Group, Institute of Computer Aided Automation, Vienna University of Technology,\\
\email{\{k\} @ auto.tuwien.ac.at}}

\maketitle

\begin{abstract}
WPA2-Personal is widely used to protect Wi-Fi networks against illicit access.
While attackers typically use GPUs to speed up the discovery of weak network passwords,
attacking random passwords is considered to quickly become infeasible with increasing
password length.
Professional attackers may thus turn to commercial high-end FPGA-based cluster solutions
to significantly increase the speed of those attacks.
Well known manufacturers such as Elcomsoft have succeeded in creating world's
fastest commercial FPGA-based WPA2 password recovery system,
but since they rely on high-performance FPGAs the costs of
these systems are well beyond the reach of amateurs.
In this paper, we present a highly optimized low-cost FPGA cluster-based WPA-2 Personal password recovery
system that can not only achieve similar performance at a cost affordable by amateurs, but in comparison
our implementation would also be more than $5$ times as fast on the original hardware.
Since the currently fastest system is not only significantly slower but proprietary as well, we believe that
we are the first to present the internals of a highly optimized and fully pipelined FPGA WPA2 password recovery system.
In addition, we evaluated our approach with respect to performance and power usage and compare it to GPU-based systems.
To assess the real-world impact of our system, we utilized the well known Wigle Wi-Fi network dataset 
to conduct a case study within the country and its border regions. Our results indicate that our system could be used to break
into each of more than $160,000$ existing Wi-Fi networks requiring $3$ days per network on our low-cost FPGA cluster in the worst case.
\end{abstract}

\keywords{FPGA, WPA2, Security, Brute Force, Attacks}

\section{Introduction}
\label{sec:Introduction}

Security in wireless Wi-Fi networks has come a long way. In comparison to wired network infrastructures, attackers are able to
easily access Wi-Fi networks if they are in the vicinity.
To protect Wi-Fi networks and the data being transfered over them, from the very beginning cryptographic protection mechanisms providing
properties such as confidentiality, integrity or authenticity have been specified in the Wi-Fi IEEE 802.11 standard documents \cite{citeulike:12556290}.
At the time WEP (Wired Equivalent Privacy) quickly turned out to be insecure allowing key recovery within minutes \cite{BreakingWEP, Bittau:2006:FNW:1130235.1130389},
manufacturers started to implement several non-standard fixes such as WEP2 or WEPplus \cite{5234856}.
Ultimately, a switch-over to WPA (Wi-Fi Protected Access) employing the RC4-based TKIP (Temporal Key Integrity Protocol) as interim solution and
to the longterm solution WPA2, in particular, has been suggested in the IEEE 802.11 standard documents \cite{citeulike:12556290}.
Since 2012, WPA has been officially deprecated in the IEEE 802.11 standard and suffers from security vulnerabilities on its own \cite{Vanhoef:2013:PVW:2484313.2484368}.
In contrast, WPA2 is FIPS 140-2 compliant \cite{NIST140-2}, much stronger and widely used to protect today's Wi-Fi infrastructures.
The WPA2-Personal variant is designed for smaller networks and uses a pre-shared key (i.e., a Wi-Fi password) to derive the necessary key material for authentication, encryption and
integrity protection. The Wi-Fi password needs to be at least 8 characters long and the key material is mainly derived through the state-of-the-art
salted key derivation function PBKDF2 (Password-Based Key Derivation Function 2) \cite{rfc2898} in combination with the SHA1 hashing algorithm \cite{rfc3174}
in HMAC configuration \cite{rfc6234}. As a result, the security of a WPA2-Personal protected Wi-Fi network heavily relies on
the quality of the password. Due to the computational complexity of the key derivation function and the use of the Wi-Fi's SSID as cryptographic salt, brute force attacks
are very hard to conduct in the presence of random passwords with increasing length.
Incurring significant costs well outside of what amateurs can afford, professional attackers can turn to commercial high-end FPGA-based cluster solutions
achieving WPA-2 password guessing speeds of 1 million guesses per second and more \cite{PicoComputing}.

In this paper, we focus on the WPA2-Personal key derivation function and low-cost FPGA cluster based attacks
that are not only affordable by professionals but by amateurs as well. Especially considering
second-hand FPGA boards that have been used for cryptocurrency mining, those boards are now available
to amateurs at low cost and can be repurposed to mount attacks on cryptographic systems.
In the first part, we use a top-down approach to present WPA2-Personal security at a high level and we subsequently
break it down to low-level SHA1 computations in high detail.
In the second part, we use a bottom-up approach to show how these computations
can be especially well addressed in hardware with FPGAs and we present how our solution can be integrated into a scalable low-cost system to
conduct WPA-2 Personal brute force attacks. We evaluate our system with respect to performance and power usage, we compare
it to results we obtained from GPUs and we conduct a real-world security evaluation case study showing the practical security
impact of our system. Specifically, the contributions presented in this paper are as follows:

\begin{itemize}
 \item We present a highly optimized design and architecture of a scalable and fully pipelined FPGA implementation for efficient WPA2 brute force attacks
 that brings the performance of today's highly expensive professional systems to the low-cost FPGA boards affordable by amateurs.
 
  \item Our implementation on Kintex-7 devices indicates that on the same hardware, 
 our implementation is more than $5$ times as fast in comparison to what is currently marketed to be world's fastest
 FPGA-based WPA2 password recovery system \cite{Elcomsoft, PicoComputing}.
  
 \item We implemented and evaluated our approach on three different low-cost FPGA architectures including an actual FPGA cluster
 comprising 36 Spartan 6 LX150T devices \cite{Spartan6Family} located on a total of 9 second-hand repurposed cryptocurrency mining boards. 

 \item We evaluate our system with respect to the power consumption and performance in comparison to GPU clusters,
 showing that FPGAs can achieve comparable or higher performance with considerably less power and space requirements, allowing attackers to create
 small and easy to use clusters.

 \item To highlight the practical real-world implications, we used the Wigle WiFi network dataset \cite{Wigle} to conduct a
 case study involving more than $166,988$ distinct Wi-Fi networks in 7 countries with potentially weak default passwords.
 Our results indicate that our system could be used to break into each of those networks requiring $3$ days per network on our low-cost FPGA cluster in the worst case.
\end{itemize}

\section{State-of-the-Art and Related Work}
\label{sec:WPA2ClusterRelatedWork}

Since WPA2 is commonly used, there are several publications and projects dealing
with WPA2 security and brute force attacks in particular. However, most of them rather
focus on GPU brute force approaches and do not cover special purpose FPGA hardware, especially considering
low-cost FPGA hardware that is available to amateurs as well.
For instance in \cite{JMEDSWPA_WPA2_Password_Security_Testing_using_Graphics_Processing_Units}, Visan
covers typical CPU and GPU accelerated password recovery approaches with state-of-the-art tools like
aircrack-ng\footnote{\url{http://www.aircrack-ng.org}} or Pyrit\footnote{\url{https://code.google.com/p/pyrit}}.
He considers a time-memory tradeoff usable for frequent
Wi-Fi SSIDs and provides a performance overview of common GPUs and GPU cluster configurations.
In that respect, oclHashcat\footnote{\url{http://hashcat.net/oclhashcat}} and the commercial Wireless Security Auditor software\footnote{\url{https://www.elcomsoft.com/ewsa.html}}
need to be mentioned as well which are both password recovery frameworks with GPU acceleration and WPA2 support.
Unlike these GPU-based approaches, our system comprises of a highly optimized and scalable FPGA implementation
allowing higher performance at lower costs and power consumption in comparison.
In~ \cite{JohRog15A}, Johnson et al. present an FPGA architecture for the recovery of WPA and WPA2 keys.
Although WPA support is mentioned, their implementation seems to support WPA2 only which is
comparable to our system. However, while our implementation features multiple fully pipelined and heavily
optimized cores for maximum performance, Johnson et al. only present a straight-forward sequential design
leading to a significantly less performance in comparison.
In~ \cite{HiPerfCOPACOBANA}, Güneysu et al.~present the RIVYERA and COPACOBANA high-performance FPGA
cluster systems for cryptanalysis. They provide details on exhaustive key search attacks for cryptographic
algorithms such as DES, Hitag2 or Keeloq and have a larger cluster configuration than we had available for our tests.
Yet, in contrast to our work, they do not cover WPA2 or exhaustive key search attacks on WPA2 in their work.
As a result, it would be highly interesting to evaluate our FPGA implementation on their machines.
Finally, Elcomsoft's commercial Distributed Password Recovery\footnote{\url{https://www.elcomsoft.com/edpr.html}} software
needs to be mentioned due to its support for WPA2 key recovery attacks on FPGA clusters \cite{Elcomsoft, PicoComputing}
and its claim to be world's fastest FPGA-based password cracking solution \cite{ElcomsoftFastest}.
Although there is practically no publicly available information on the internals of their WPA2 implementation,
in \cite{PicoComputing} performance data are provided. In contrast to their work, we do not only disclose
our design, architecture and optimizations of our FPGA implementation, but we also claim that on the same professional FPGA hardware
our implementation would be more than $5$ times as fast. In comparison to the professional system, our system can 
achieve similar speeds on the low-cost repurposed cryptocurrently mining hardware that is available to many amateurs.

\section{WPA2-Personal Handshake}
\label{sec:WPA2ClusterWPA2}

\begin{figure}[htbp]
\centering
  \includegraphics[width=0.65\textwidth]{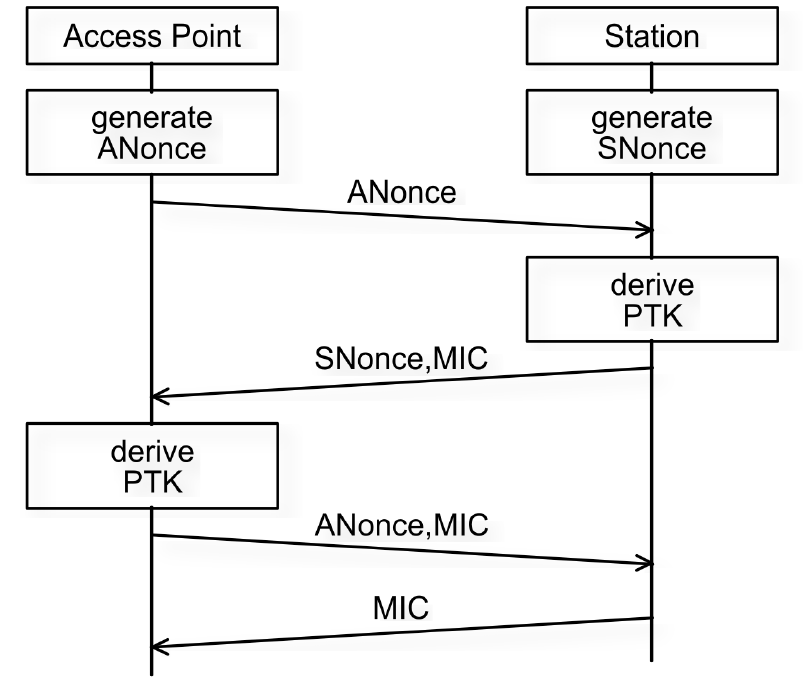}
\caption{WPA2-Personal 4-Way Handshake}
\label{fig:WPA2ClusterWPA2Handshake}
\end{figure}

Whenever a Wi-Fi client (Station) would like to connect to an Access Point, there are 802.11 management frames involved \cite{citeulike:12556290}.
For instance in non-hidden Wi-Fi networks, the Access Point typically transmits beacon frames
to advertise the network. In order to connect, a Station sends a probe request to determine network capabilities 
such as supported rates or vendor specific information. After that, in WPA2-Personal protected networks, 
Station and Access Point mutually authenticate against each other with the 4-way handshake 
depicted in Fig. \ref{fig:WPA2ClusterWPA2Handshake}.

To start the mutual authentication process, the Access Point generates
a 32 byte random \texttt{ANonce} and sends it to the Station.
Similarly, the Station generates a 32 byte random \texttt{SNonce}
and uses both nonces as well as the secret password to derive the PMK (Pairwise Master Key) and the
Pairwise Transient Key (PTK) with the help of the WPA2-Personal key derivation functions described in the following Section \ref{subsec:WPA2ClusterWPA2KeyDerivation}.
The nonces ensure that the handshake cannot by replayed by an attacker at a later time.
Afterwards, the Station sends the \texttt{SNonce} back to the Access Point
and utilizes the \texttt{PTK} truncated to the first $128$ bits (denoted Key Confirmation Key - \texttt{KCK}) to compute a Message Integrity Code (MIC) over the packet data.
At this point, the Access Point can already compare the received MIC with the computed one
to validate that the Station is authentic and has knowledge of the secret password.
In order to prove to the Station that the Access Point knows the secret password as well,
the Station sends a message including \texttt{ANonce}
and the corresponding \texttt{MIC} code. Since the Station can only compute the correct MIC code
if it knows the necessary \texttt{PTK}, the Access Point can use this information for authentication.
If the authentication was successful, the Station completes the handshake by sending a usually empty, but signed
(\texttt{MIC}) message back to the Access Point.
The Station can now associate to the Access Point and take part in the Wi-Fi network.

\subsection{Key Derivation}
\label{subsec:WPA2ClusterWPA2KeyDerivation}

\begin{figure}[htbp]
\centering
  \includegraphics[width=0.45\textwidth]{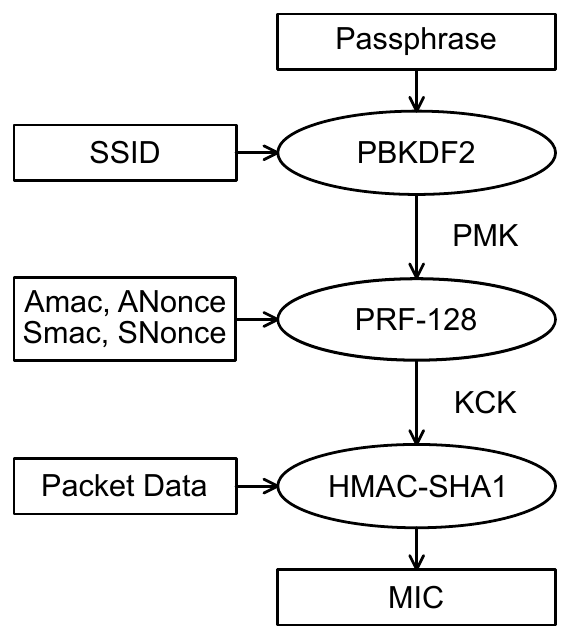}
\caption{WPA2-Personal Key Derivation Function}
\label{fig:WPA2ClusterKeyDerivation}
\end{figure}

In order to compute the \texttt{PTK} and its truncated variant (denoted the \texttt{KCK}) required to
compute the \texttt{MIC} integrity code for provided packet data, the key derivation algorithm visible in Fig. \ref{fig:WPA2ClusterKeyDerivation} is utilized in WPA2-Personal.
It uses the pre-shared secret key (i.e., the Wi-Fi network passphrase) and provided network information such as the SSID (i.e., the Wi-Fi network name), nonces
and the MAC addresses as inputs.
To achieve a high level of security, at least two factors need to be considered in the key derivation.
First, the key derivation algorithm needs to be collision resistant and computationally expensive.
Collision resistant denotes the property that it is hard to find two different inputs that result in the same hash output
when the hash function is applied. If the hash function is also computationally expensive, it will
take an attacker longer to compute hash outputs thereby slowing down the number of guesses he can make per second.
The longer and more complex the Wi-Fi password is, the more possible password combinations exist and the more hash computations
the attacker needs to make to find the correct password.
Second, the key derivation algorithm needs to be cryptographically salted so that, depending on the salt,
different keys are generated for the same password.
The general idea of salting is to add a random value to the message before the hash function is applied while
the salt value is stored for later use.
As a result, the same password will lead to different hash outputs since the salt value is different.
Without the use of a salt, attackers could pre-compute lookup-tables for all possible passwords and corresponding hashes.
While the size requirements of this table would grow tremendously with increasing password lengths, a practical
time/memory tradeoff can be achieved with a pre-computed \emph{rainbow table} \cite{citeulike:1855432}.
The general idea of a rainbow table is to only store a small part of the possible hash value and password combinations.
This is achieved by choosing a random password candidate as starting point and applying the hash function on it.
However, instead of storing the hash value, the key idea is to define a reduction function that uses the hash value as input to
create another valid password candidate.
This process is continued to create entire \emph{chains} where each chain ends either if the given length has been achieved or a password candidate has been created
that is already a starting point for one of the already created chains. Since only the starting point and endpoint password candidates are stored,
the storage requirements can be lowered.
Once all possible chains have been pre-computed, the attacker can start to look up the password for a given hash value by computing the
reduction function and locating the resulting password candidate in the stored endpoints of the chains in the rainbow table.
As not all password candidates are stored in the table, it might very well be the case that the candidate can not be found.
The attacker thus computes the hash output for the password candidate and applies the reduction function on the corresponding output to get another password candidate.
This computationally expensive step is repeated until the password candidate is found among the endpoint password candidates of the chains in the table.
Once found, the attacker takes the starting point password candidate and recomputes the intermediary values in the chain.
At some point the computation will result in the hash value the attacker is looking for. The password will be
the previously computed input value prior to this hashing step or the starting point password candidate itself (if it is the first hash value).
Prior to computing a rainbow table, an attacker can thus freely choose the time/memory tradeoff between the required lookup time and the required storage space for the table.
For more information, we would like to point to Martin Hellman's original paper \cite{citeulike:1855432}.
To mitigate these threats, WPA2-Personal relies on the salted PBKDF2 \cite{rfc2898} key derivation function.
In the following, we describe the key derivation process in detail and closely focus on the required computational
effort due to its impact on FPGA implementations and the achievable password guessing speed.

\subsection{Breaking it down to SHA1 Computations}
\label{subsec:WPA2ClusterWPA2SHA1}

\begin{figure}[htbp]
\centering
  \includegraphics[scale=0.80]{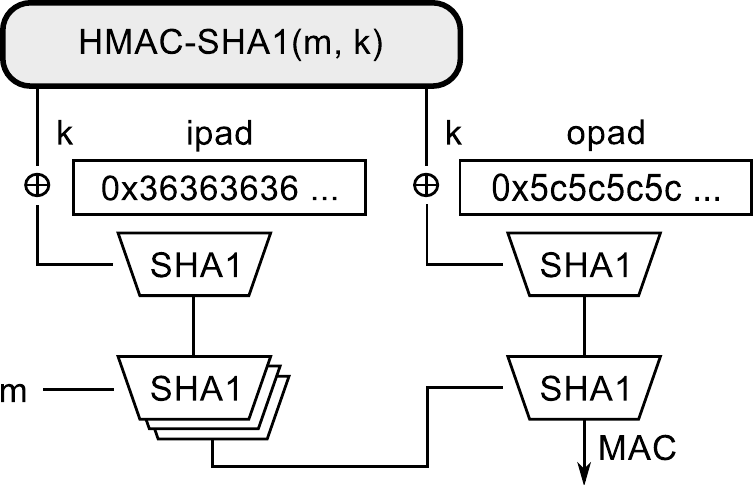}
\caption{PBKDF2 core with SHA1 rounds in HMAC construction}
\label{fig:WPA2ClusterHMAC}
\end{figure}

Internally, the PBKDF2 key derivation function employed in WPA2-Personal utilizes $4,096$ iterations of the well known HMAC construction with the SHA1 cryptographic hash algorithm at its core
to obtain $160$ bit hash outputs (Fig. \ref{fig:WPA2ClusterHMAC}).
Since the WPA2 Pairwise Master Key \texttt{PMK} needs to be $256$ bits long, two PBKDF2 rounds are necessary.
Their output is concatenated, but from the second iteration the output is truncated to $96$ bits to achieve a $256$ bit result.
In both PBKDF2 iterations the secret password is used as key while the SSID of the Wi-Fi network concatenated with a $32$ bit
counter value serves as input. In the first iteration, the counter value is one while in the second iteration it is two.
Consequently within both PBKDF2 iterations, there are $8,192$ HMAC-SHA1 iterations required to compute the \texttt{PMK} from the secret password and the network's SSID.
With regard to the HMAC internals, Fig. \ref{fig:WPA2ClusterHMAC} shows that a number of SHA1 iterations are necessary to obtain the MAC (Message Authentication Code).
In general to compute the SHA1 hash digest of a message, the first SHA1 iteration is computed by using the initial SHA1 state and hashing the first part of the message.
Depending on the length of the message, additional iterations might be necessary whereupon the previous SHA1 state output is used as state input for the next iteration.
Once the full message has been hashed, SHA1 finalization needs to be applied by appending a '1' bit and the length of the message to the message itself and filling up the rest of the 
$512$ bit SHA1 input block with '0' padding bytes.
For WPA2-Personal key derivation, in the first PBKDF2 round the xor-transformation is applied on the password and the inner pad \texttt{ipad}. The result is a $512$ bit block serving as input
to the SHA1 hash function in initial state. The output is the HMAC inner state. Since the SSID may be no longer than $32$ bytes, the hashing of the SSID and the $32$ bit PBKDF2 round
counter can be done together with the SHA1 finalization so that only one SHA1 iteration is necessary.

In the next step, the outer HMAC state is computed by hashing the xor of the password and the outer pad \texttt{opad}. Afterwards, the
previously finalized $160$ bit digest is hashed and finalized with the outer state. At this point the MAC is ready.
The second PBKDF2 iteration is computed in the same way with the difference that the round counter value is set to two instead of one.
Since the password does not change during PBKDF2 iterations, the inner and outer HMAC states stay the same
allowing us to use cached states instead of having to compute the states again.
With that optimization in mind, it is required to compute at least $2 + 4,096*2$ SHA1 iterations for the first PBKDF2 round
and $4,096*2$ SHA1 iterations for the second round (i.e., $16,386$ SHA1 iterations in total) to obtain the \texttt{PMK}.
This computational effort, the use of the SSID as salt for key derivation and the security of the innermost SHA1 cryptographic hash function
are three of the main reasons why WPA2-Personal key derivation is considered to be very strong against typical exhaustive key search attacks.

Once the \texttt{PMK} is available, the \texttt{KCK} is derived by applying a $128$ bit Pseudo Random Function (PRF).
Internally, it just uses HMAC-SHA1 again with the \texttt{PMK} as key. The hashed message is made up of the string ``Pairwise key expansion'', a terminating zero byte,
an arithmetically sorted tuple of the Access Point and Station addresses as well as another sorted tuple of their nonces (i.e., \texttt{ANonce} and \texttt{SNonce})
including a finalizing zero byte. The \texttt{PTK} is the resulting MAC and it is truncated to the first $128$ bits to obtain the \texttt{KCK}.
If the \texttt{PMK} is available, the computation
of the \texttt{KCK} takes $5$ SHA1 iterations as due to the length of the \texttt{PMK} the finalization of the inner HMAC state can not be combined with
the hashing of the \texttt{PMK}.

Whenever Access Point or Station would like to compute a \texttt{MIC}, they can do so by utilizing HMAC-SHA1 on the message with \texttt{KCK} as key.
The result of the computation truncated to the first $128$ bits is the \texttt{MIC}.
The computational effort depends on the length of the message.
However, considering the messages from the 4-way WPA2-Personal handshake, a total of $5$ SHA1 iterations
is required to compute the \texttt{MIC} since, similar to the \texttt{KCK} computation, the finalization of the inner HMAC state
requires one additional iteration.

\subsection{SHA1 Internals}
\label{subsec:WPA2ClusterSHA1Internals}

Due to the high number of required SHA1 computations, it is essential
to increase their speed as much as possible.
To compute a SHA1 hash, a number of computational steps is necessary.
Due to the high impact on our FPGA implementation, we provide a detailed overview
of SHA1 internals.
SHA1 \cite {NIST180-2} works on $512$ bit chunks and produces a $160$ bit hash digest when finished.
If the message length is less than $512$ bit, padding bits are used. For SHA1 finalization, a '1' bit, the padding bits (if necessary) and a $64$ bit length field are
appended.
SHA1 has $80$ internal rounds (denoted \texttt{t}) and requires a separate message working schedule \texttt{$W_{t}$} as well as a constant \texttt{$K_{t}$} for each of them.
In the pre-processing step, the message working schedule \texttt{$W_{t}$} is computed as follows:

\begin{eqfloat}
        \begin{equation*}
        \resizebox{0.8\hsize}{!}{$W_t = \left\{
  \begin{array}{l r}
    M_t & 0 \leq t \leq 15\\
    rol(W_{t-3} \oplus W_{t-8} \oplus W_{t-14} \oplus W_{t-16}, 1) & 16 \leq t \leq 79\\
  \end{array} \right.
$}
        \end{equation*}
\end{eqfloat}

The schedule \texttt{$W_{0} \ldots W_{15}$} is the message broken up into $16$ words with $32$ bit length each.
For the remaining $64$ words, message expansion is used by applying the xor-operation on previous schedules and rotating the result
one time to the left. The constants \texttt{$K_{t}$} for the rounds comprise of a set of $4$ words:

\begin{eqfloat}
        \begin{equation*}
  K_t = \left\{ 
  \begin{array}{l r}
    \text{5a827999} &  0 \leq t \leq 19\\
    \text{6ed9eba1} & 20 \leq t \leq 39\\
    \text{8f1bbcdc} & 40 \leq t \leq 59\\
    \text{ca62c1d6} & 60 \leq t \leq 79\\
  \end{array} \right.
        \end{equation*}
\end{eqfloat}

\begin{figure}[htbp]
\centering
  \includegraphics[scale=0.8]{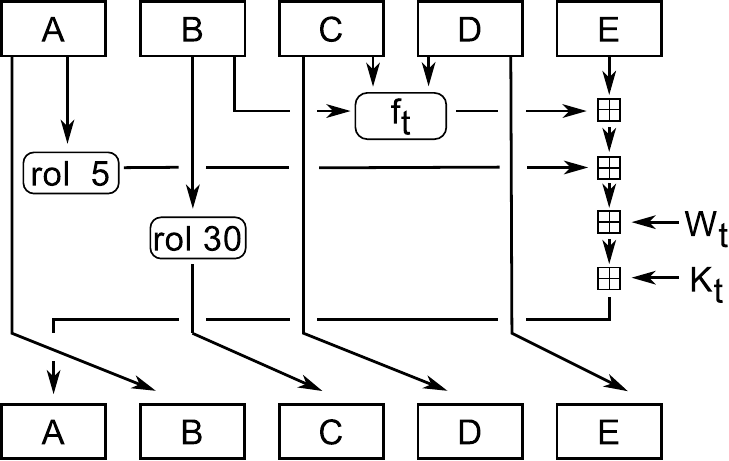}
\caption{SHA1 Compression Round}
\label{fig:WPA2ClusterSHA1Compression}
\end{figure}

After precomputation, the $80$ SHA1 rounds are performed. Each round is based on the compression
round visible in Fig. \ref{fig:WPA2ClusterSHA1Compression} and works on five $32$ bit words denoted \texttt{A} to \texttt{E} where \texttt{rol n} denotes a rotate left by \texttt{n} operation and the
$\boxplus$ operator denotes an unsigned 32 bit addition.
In the initial iteration, a constant initialization vector \texttt{H0} to \texttt{H4} is used as input for \texttt{A} to \texttt{E}.
The difference between the rounds is the function \texttt{$f_t$} defined as follows:

\begin{eqfloat}
        \begin{equation*}
  f_t = \left\{ 
  \begin{array}{l r}
    (x \land y) \oplus (\lnot x \land z) 		&   0 \leq t \leq 19\\
    x \oplus y \oplus z 				&  20 \leq t \leq 39\\
    (x \land y) \oplus (x \land z) \oplus (y \land z)	&  40 \leq t \leq 59\\
    x \oplus y \oplus z 				&  60 \leq t \leq 79\\
  \end{array} \right.
        \end{equation*}
\end{eqfloat}

After each round, the resulting words \texttt{A} to \texttt{E} are fed back as input to the next round.
Once all $80$ rounds have been computed, the resulting words are added to the initialization vector \texttt{H0} to \texttt{H4}
and the concatenated result is the resulting hash digest. Subsequent SHA1 computations are computed in the same way except that
instead of the initialization vector the hash digest from the previous block is used.

\subsection{Attacking the 4-Way Handshake}
\label{subsec:WPA2ClusterWPA2Attack}

If an attacker would like to determine the secret WPA2-Personal password, a 4-way WPA2-Personal handshake
between a Station and the Access Point needs to be obtained first.
This can either be done passively or with the help of an active de-authentication attack where the attacker
spoofs the source address of the Access Point and sends de-authentication frames to the Station.
Since those frames are not authenticated, the Station will falsely believe that the de-authentication request came from the genuine Access Point
and will follow the request. However at a later time, it will re-authentication and thus give the attacker the opportunity to intercept
the handshake. As soon as the attacker has the handshake, passwords can be guessed offline by deriving the key material for the \texttt{PMK} and the \texttt{KCK}
and computing the \texttt{MIC} for one of the observed packets in the handshake.
If the observed \texttt{MIC} is the same as the computed \texttt{MIC} for a password candidate, the attacker has found the correct secret password for the network.
However, since a WPA2-Personal password needs to have a minimum length of $8$ characters and for each password candidate a total of at
least $16,386 + 5 + 5 = 16,396$ SHA1 iterations are necessary to compute the corresponding \texttt{MIC} over a handshake packet,
exhaustive password guessing attacks are considered to be increasingly infeasible with higher password complexity and length.
In the subsequent chapters, we show that the high computational effort can be addressed with special purpose FPGA hardware
so that a high number of real-world WPA2-Personal protected networks with random passwords can be broken into within days.

\section{FPGA Implementation}
\label{sec:WPA2ClusterWPA2ClusterFPGA}

Implementing an algorithm on FPGAs is significantly different from software implementations. The FPGA comprises of
different building blocks such as RAM or configurable logic blocks containing LUTs (Look Up Tables), Flip-Flops, dedicated arithmetic logic or shift registers.
Initially, the inputs and outputs of all these building blocks are unconnected. During FPGA configuration,
a bit-stream is uploaded to the configuration memory of the FPGA that subsequently sets up the interconnections using switch boxes.

Assuming at least some familiarity with the FPGA design, in the first step the designer utilizes a hardware description language (HDL) such as VHDL to describe the design.
Based on the design, a synthesis tool creates a netlist that transfers the design to a set of interconnected high-level logic components.
The netlist can thus be seen as a high-level schematic comprising the information which components are interconnected to each other through signals.
In the next step, the components in the netlist are mapped to the building blocks of the specific targeted FPGA device such as LUTs, dedicated shift registers or memories.
Afterwards, similar to the work that needs to be done when designing a printed circuit board from a schematic, the components within the implementation need to be placed
within the FPGA and the interconnects between those blocks need to be routed.
The mapping, placing and routing steps are especially critical as often millions of interconnects need to be made
and the signal run time for each of them needs to stay within specification. If only one signal requires a longer time
from one register to the next one, the maximum clock frequency of the whole FPGA implementation will decrease
to the clock frequency supported by the slowest path (i.e., the \emph{critical path}).
On the other hand the placement of the components is of paramount importance as well. If two components are placed non ideally and too far apart, their interconnects need not only be routed
across a large area, but the time it takes for a signal to be transferred will also increase significantly thereby lowering the
maximum clock speed of the entire implementation.
Creating and especially optimizing high-speed FPGA implementations is thus highly challenging as apart from the logic design
physical constraints such as the mapping, the signal routing or the electrical loads of signals need to be addressed as well.
Since for large designs these steps often require multiple hours of design tool run time, performing optimizations can be hard
as each design adaptation often requires another full design tool run. Only after the entire design flow has completed, the designer can
get an impression whether the optimization was beneficial or not.
However, the advantage is that depending on the requirements of an algorithm such as memory usage or the number of logic cells,
it is often possible to achieve tremendous speedups, high scalability and lower power usage by efficiently using the resources of the FPGA
and carefully optimizing the outcomes of the design steps from netlist generation to the final generation of the FPGA bit stream file.
Considering the SHA1 internals described in Section \ref{subsec:WPA2ClusterSHA1Internals}, the algorithm is especially well suited
for FPGA implementation due to the following reasons:

\begin{enumerate}
 \item The algorithm has practically no memory requirements.
 \item The rotate and shift operations utilized in SHA1 can be realized through FPGA interconnects with minimal time delay
 \item Algebraic logic functions (xor, and, or, not, etc.) require minimal effort and can efficiently utilize the FPGAs LUTs
\end{enumerate}

The most expensive operation are SHA1's additions due to the long carry chain between the adders. To implement the algorithm, a surrounding state machine
is required to control which inputs should be supplied to the logic in different rounds.
Considering that SHA1 has $80$ rounds and we would like to achieve maximum performance, there are two design options:
Either the SHA1 algorithm is implemented sequentially or in a fully pipelined way.

The advantage of a sequential implementation is that the FPGA can be completely filled up with relatively small SHA1 cores. However, the disadvantage
is that each of those cores would require its own state machine which takes up a significant amount of space.
In comparison, a fully pipelined implementation does not require an internal state machine as each of the SHA1 rounds
is implemented in its own logic block. While this is a significant advantage enabling parallel processing, the drawback is that a fully pipelined implementation has
much higher space and routing requirements. When using multiple cores (each containing a full pipeline), only an integer number of cores can be placed so that a significant amount of
unused space might be left on the FPGA. In our implementation, we also experimented with filling up this space with sequential cores but refrained from it
due to the negative effect on the overall design complexity and the lower achievable clock speeds.

Due to the typically higher performance that can be achieved through pipelining and the property that we get
one full SHA1 computation output per clock cycle per core, we targeted a heavily optimized and fully-pipelined approach.
However, while pipelining alone has a considerable performance impact in comparison to a sequential approach, the key
of obtaining maximum design performance are the optimizations.

\subsection{FPGA Design}
\label{subsec:WPA2ClusterFPGADesign}

\begin{figure}[htbp]
\centering
  \includegraphics[scale=0.9]{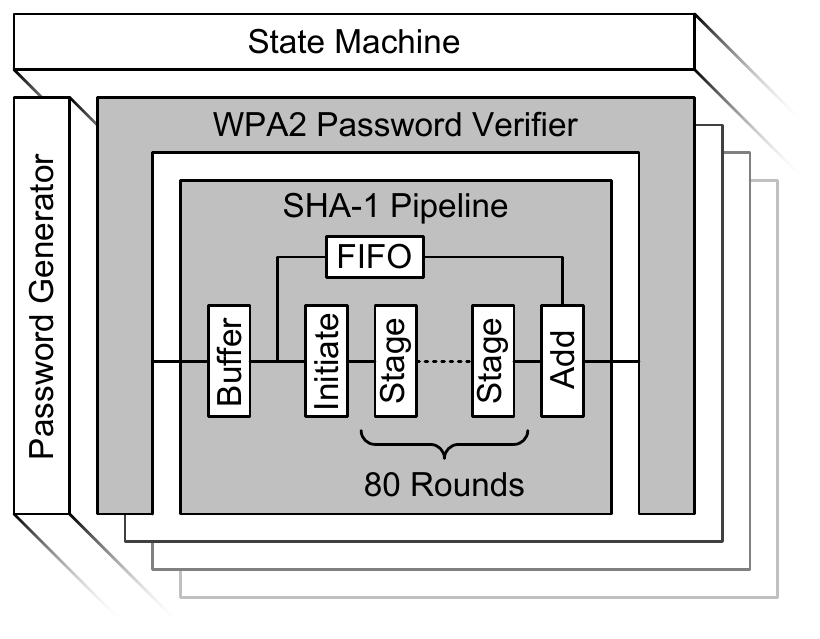}
\caption{FPGA Design Overview}
\label{fig:WPA2ClusterFPGADesign}
\end{figure}

Our overall FPGA design is illustrated in Fig. \ref{fig:WPA2ClusterFPGADesign} and has the following components:
A shared password generator, a global brute force search state machine and an FPGA device specific number of
brute force cores, each comprising a WPA2-Personal state machine with password verifier and a SHA1 pipeline.

\subsubsection{Password Generator}
\label{subsubsec:WPA2ClusterPasswordGenerator}

\begin{figure}[htbp]
\centering
  \includegraphics[width=0.4\textwidth]{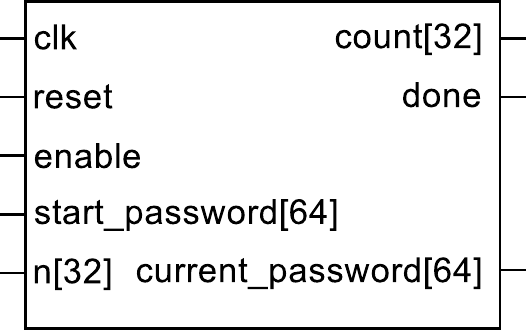}
\caption{Password Generator Block}
\label{fig:WPA2ClusterPWGen}
\end{figure}

The password generator (Fig. \ref{fig:WPA2ClusterPWGen}) is realized as a fast counter.
Whenever the FPGA is idle, it can accept a new
working block comprising of all necessary data including the actual start password (\texttt{start\_password})
and how many passwords (\texttt{n}) should be tested.
Initially starting at the start password, whenever the password generator is enabled (\texttt{enable})
it will output a new password (\texttt{current\_password})
and the current password number (\texttt{count})
in each clock cycle.
In case no more passwords can be fed into the brute force cores, the generator can be paused at any time by
disabling the \texttt{enable} input. Ultimately, it will output new passwords until \texttt{n} passwords have been
reached and assert the \texttt{done} signal to indicate that all passwords within the current working block have been generated.

During the optimizations of our cryptographic cores in the design, at some point
the long carry chain in the password counter became the clock speed limiting critical path. 
We were able to address the issue by parallelizing the counter and implementing
the password carry with static multiplexers outside the sequential logic block.
The sequential logic block can be seen as typical register transfer logic (RTL).
With the clock signal, the old counter value is fetched from the source register, increased and
finally output to the destination register. The path in between accounts for the delay.
Since we need to have a carry overflow at the last valid password character (e.g., 'Z') we need
a set of multiplexers that eventually reset the characters at each position of the password string.
However, if this multiplexer based reset logic is within the sequential path it will also increase
the time delay. By statically implementing the reset logic outside this sequential path we were able
to balance the overall worst-case delays and achieved a password counter implementation
that no longer accounted for the critical path in our overall design.
Another password generator optimization approach we considered is utilizing multiple clock domains.
The general idea is that the overall design naturally spends most of its time computing SHA1 iterations.
At that time the password generator is disabled. We could thus use a less critical slower clock
to generate the passwords and output them to clock synchronizing FIFO buffers directly placed next to the input of the SHA1 pipelines.
As soon as a SHA1 pipeline requires a new password input, it can utilize its fast clock to drain the FIFO buffer which would in turn
enable the password generator to refill the corresponding buffer at its slower clock.
The advantages of this approach would be the following: First, the complexity of the password generator design
can be further increased without negatively impacting the critical path. However second, the big advantage
is the routing of the bus signals from the password generator to all the cores. Considering that the password
generator is located at the center of the design and the passwords need to be distributed across the
entire FPGA to all brute force cores, there is a significant impact on the time-driven routing complexity and the
interconnect delays that negatively impact the maximum clock speed of the overall design.
By leveraging a slower clock, the passwords would be already located in the FIFO buffers next to the SHA1 pipelines of
each core but they could still be read with the fast clock the SHA1 pipelines are operating on.
However, since with our previously mentioned password generator optimization the critical path was no longer
within the password generator domain, we did not implemented the approach. It will be covered in future work.

\subsubsection{Global Brute Force State Machine}
\label{subsubsec:WPA2ClusterBruteforceStateMachine}

The task of the global brute force state machine is to constantly supply all brute force cores
with new password candidates and check whether one of them found the correct password.
Due to the insignificant speed impact and the advantage of lower design complexity we chose an iterative approach.
Since our SHA1 pipeline comprises of $83$ stages, we can concurrently test $83$ passwords per brute force core.
With our iterative approach, we enable the password generator and consecutively fill all brute force cores with passwords.
Once all cores have been filled, the password generator is paused and we iteratively wait until all cores
have completed. At that point, the password filling process is restarted.
If a core finds the correct password or the password generator has reached the last password, the state
machine jumps into the idle state and can accept the next working block.
The penalty for this iterative approach is $83$ clock cycles per core since once a brute force core
has completed, we could immediately fill it with a new password.
However, in comparison to the long run time of each core the impact is insignificant.

\subsubsection{WPA2-Personal State Machine with Password Verifier}
\label{subsubsec:WPA2ClusterWPA2StateMachine}

\begin{figure}[htbp]
\centering
  \includegraphics[width=0.7\textwidth]{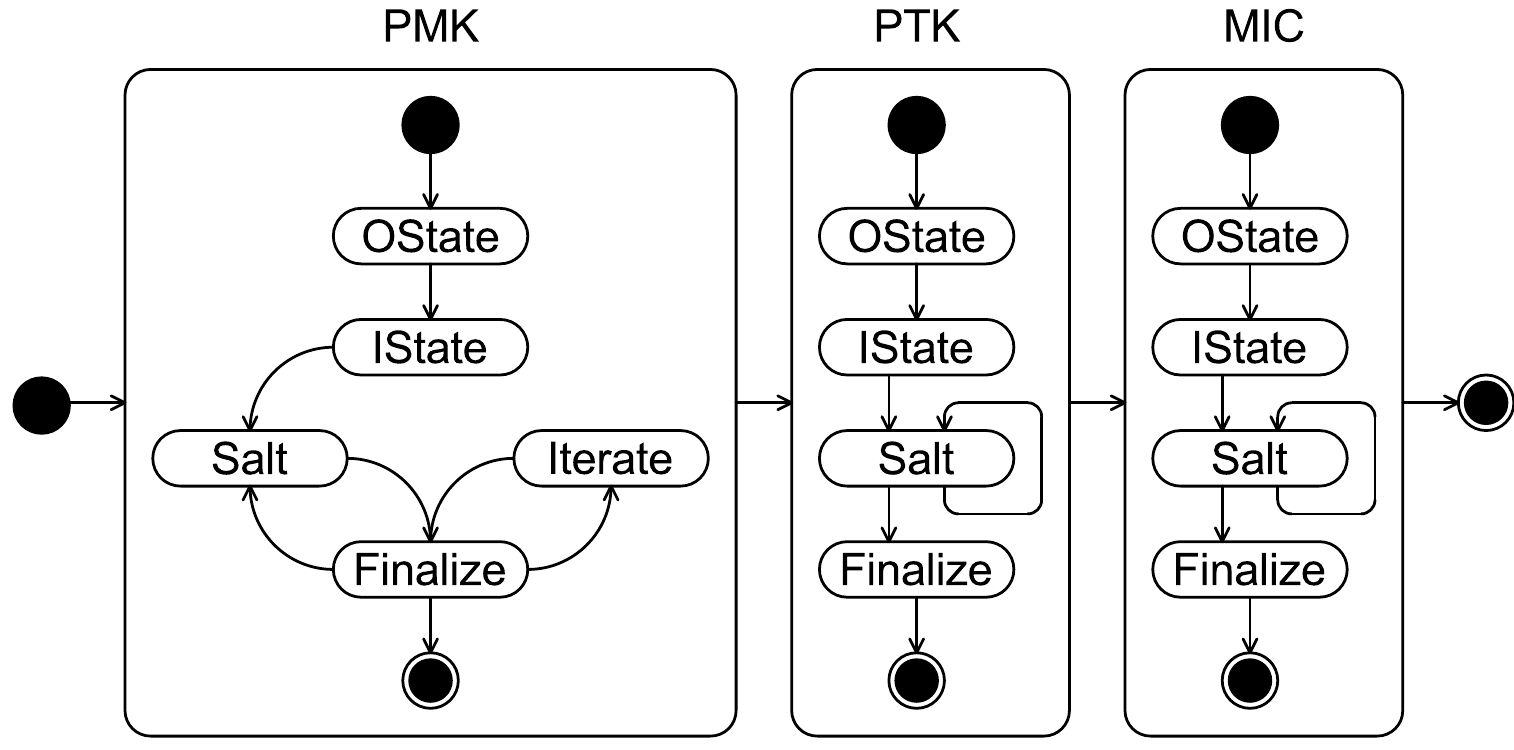}
\caption{WPA2-Personal FPGA States}
\label{fig:WPA2ClusterWPA2states}
\end{figure}

Each brute force core has a WPA-2 Personal state machine with a password verifier.
It is the most complex state machine in the overall design. Its task is
to compute the \texttt{MIC} code for each password candidate with the help of
the SHA1 pipeline in its center. Each computed \texttt{MIC} is compared with the
\texttt{MIC} from the WPA2-Personal 4-way handshake to determine whether the password
candidate was correct or not. Figure \ref{fig:WPA2ClusterWPA2states} shows all necessary states and state
transitions.

The state machine is divided into three WPA2-Personal key derivation phases: PMK computation (1), PTK computation (2), and MIC computation (3).
The computation of the PMK has the highest computation effort due to the $2$ PBKDF2 rounds with $4,096$ iterations requiring $16,386$ SHA1 iterations in total.
Initially, $83$ password candidates and the network's \texttt{SSID} are fed into the SHA1 pipeline to compute the corresponding HMAC outer and inner states (\texttt{OState} and \texttt{IState}).
Since these states do not change over the PBKDF2 iterations, the HMAC state computation needs to be done only once.
In the first PBKDF2 round, the \texttt{SSID} and the PBKDF2 round counter ($1$) are used as salt. After that, there are $4,095$ more iterations in which the digest output
is used as input. At that point, the second PBKDF2 round is computed by first computing the salt with an increased round counter value ($2$) and subsequently
performing $4,095$ iterations to obtain the \texttt{PMK}.

\subsubsection{SHA1 Pipeline}
\label{subsubsec:WPA2ClusterSHA1Pipeline}

In each brute force core, the SHA1 pipeline occupies a large amount of space due to the high number of pipeline stages.
While SHA1 has $80$ rounds and a fully pipelined implementation would thus have an equal number of pipeline stages,
we heavily optimized our pipeline to allow higher clock frequencies and consequently achieve more performance.
The SHA1 pipeline is the key limiting factor of how fast our password guessing attacks can be conducted.
Within the brute force cores, each of our SHA1 pipelines has $83$ stages due to the optimizations we performed.
Each core can thus compute $83$ password candidates in parallel.
The optimization approaches we applied are described in the following:

The first stage of the SHA1 pipeline is a buffer stage so that the delays of the different
input logic blocks within the WPA-2 Personal state machine are not added to the pipeline's input logic
and thereby does not increase the overall time delay of the critical path.
The second stage denoted 'Initiate' is an optimization of the $4$ required (expensive) additions in the \texttt{E} word of each SHA1 round. Instead
of having all $4$ additions in one stage, the structure of the SHA1 algorithm allows to pre-compute the output of the \texttt{f} function.
The addition of the \texttt{E} word with the output of \texttt{f} and the key \texttt{$K_t$} enabled
us to split up the required $4$ sequential additions into two rounds with $2$ additions, thereby significantly improving
the maximum clock speed.
Since the expansion steps for the message working schedule \texttt{$W_t$} require only a small amount of logic,
another optimization is to do multiple message expansion steps in a single pipeline stage so that
it is not needed in the following few stages. As a result, the source data is not accessed in each stage and
shift register inference is boosted causing lower flip-flop fan-out as well as less power usage and lower area requirements.
Another approach we took is the pipeline stage denoted 'Add' after the SHA1 rounds. After the last SHA1 round, the resulting digest
is added either to the constant initialization vector \texttt{H0} (first iteration) or to the previous digest for subsequent iterations.
Due to these expensive additions, the design performance can be improved if they are carried out in a separate pipeline stage.
Instead of forwarding the initial digest through all stages to the final addition stage,
we leverage a FIFO-based delay line utilizing the FPGAs Block-RAM resources. This avoids excessive interconnect routing
through all stages and thus makes the design smaller, reduces the number of critical paths and allows us
to achieve higher clock frequencies more easily.

\subsubsection{Additional FPGA Design Optimizations}
\label{subsubsec:WPA2ClusterFPGAOptimizations}

In the WPA2-Personal state machine, we directly use the output from the password generator and
compute the HMAC \texttt{OState} state first. At the same time, we store the password candidates in a Block-RAM buffer for later \texttt{IState} computation.
After that, we no longer work with the passwords but use password offsets instead.
The result is a lower design density as no more additional interconnects are required for the password in later stages.
A similar approach is used to avoid excessive interconnects and design density. Instead
of having large buses, we either use Block-RAMs directly or form RAM-based delay lines to keep the \texttt{IState} and \texttt{OState} states
as well as the computed \texttt{PMKs} and \texttt{PTKs} in memory.
Instead of one large WPA2-Personal state multiplexer directly controlling all SHA1 pipeline inputs and outputs, we
make use of several smaller and less complex multiplexers. Once again, this reduces overall design complexity and allows us to achieve higher clock speeds more easily.
The top-level design needs to communicate with the outside world. Each time a new working block is added, all necessary Wi-Fi and WPA2-Personal data
needs to be transferred and subsequently forwarded to all brute force cores. The result is a very broad bus spreading all over the FPGA design and causing
severe design congestion. Since in our design only the password candidates and the \texttt{SSID} are required early within the WPA2-Personal state machine, we transfer
the rest of the data over a small 16 bit bus leveraging inferred shift registers. This significantly reduces the complexity of the
interconnects between the shared global state machine and the brute force cores across the FPGA.
To lower the amount of input and output data exchanged with the outside world, we use a minimized Wi-Fi and WPA2-Personal data set
that only includes the variable data fields from the captured handshake. All other data is not only fixed within the FPGA, but also kept locally in the cores. In addition,
the FPGA does not output the correct password, but a numeric offset from the start password instead.

To avoid design congestion and to push the design to the highest clock speed possible, we make use of custom parameters within the Xilinx design tools
for synthesis, mapping and routing such as the minimum inferred shift register size, register balancing or the number of cost tables. In addition, we use floor planning
to support the mapper, placer and router in achieving higher clock rates. Floor planning is important to place critical components requiring a fast interconnect in between next to each other.
In general, we were able to obtain the highest speed improvements by utilizing a star like topography: The password generator is distributed over the very center of the FPGA and
the brute force cores are surrounding it. In addition we also used floor planning to avoid the placement of time critical components in FPGA areas that are hard to reach through interconnects.
Especially considering the low cost Xilinx Spartan-6 and Artix-7 FPGA devices, we could identify major regions that can not be used to place components or interconnects. Consequently, we carefully
placed critical components like the SHA1 pipelines in a way that those regions do not negatively impact the routing delay.
In our FPGA implementations, we use a slow clock for communication with the outside world and a fast clock for computation at the same time. In our Spartan-6
implementation, the speed of the fast clock can be adjusted dynamically during runtime by programming the clock multiplier. In contrast, our Artix-7 implementation
includes an automatic clock scaling mechanism to adjust the fast clock frequency with the device core temperature. Both approaches allow the FPGA design to run at high speeds
without the danger of overheating.

\subsection{Overall System Design}
\label{subsec:WPA2ClusterSystemDesign}

\begin{figure}[htbp]
\centering
  \includegraphics[width=0.7\textwidth]{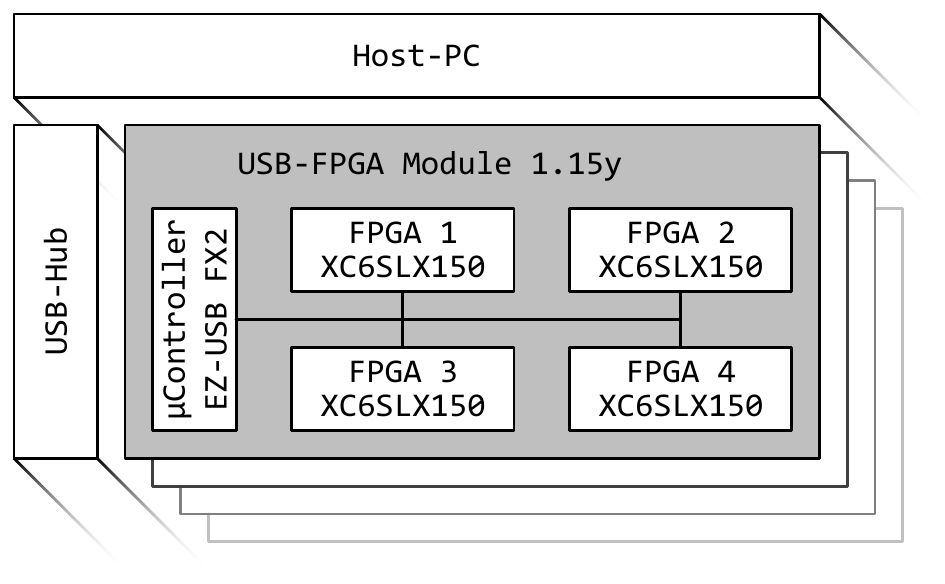}
\caption{System Overview (Spartan 6 System)}
\label{fig:WPA2ClusterSystemOverview}
\end{figure}

We implemented and practically evaluated our system on older model Xilinx Spartan-6 
as well as on newer model Xilinx Artix-7 
FPGAs. The Spartan-6 FPGAs are located on low-cost repurposed cryptocurrency mining
boards.
For comparison purposes, we created a full implementation
for the more expensive Xilinx Kintex-7 XC7K410T FPGA as well, but could not practically test it since we
did not have one of these FPGAs at hand.
The overall system design for the Spartan-6 FPGAs is visible in Fig. \ref{fig:WPA2ClusterSystemOverview}
and based on ZTEX \cite{ZTEX} FPGA boards.
The Artix-7 design is similar but has only one XC7A200T FPGA on the board.

The system comprises of a PC with a host software and several FPGA boards
connected via the USB 2.0 high-speed interface. Each FPGA board has a fast
EZ-USB FX2 micro-controller with custom firmware to interface with the FPGAs.

Our custom host software comprises $\sim$2k lines of Java code and utilizes the ZTEX SDK
to allow easy communication with the micro-controller and the FPGAs.
The host software accepts a configuration file that includes all necessary Wi-Fi and WPA-2 Personal handshake data.
At startup, it enumerates all connected FPGA boards, uploads the micro-controller firmware if necessary and 
configures the FPGAs with our bit stream.
The software makes use of several threads. Apart from the main program, there is a thread
to generate password working blocks for the FPGAs and additional threads for each FPGA board.
The password working blocks are kept in a pool with constant size. The device threads can supply working blocks to FPGAs and mark them
as being processed. If an FPGA has finished a block, it is removed from the pool and the generator automatically creates a new working block.
If for some reason an FPGA fails, the block sent to the FPGA is still in the pool and just needs to be unmarked so that the next
free FPGA can process it instead.

The micro-controller firmware comprises $\sim$1k lines of C code and is responsible for USB communication with the host
and communication with the FPGAs. Each FPGA has an 8 bit write and an 8 bit read bus in addition to
read and write clocks, a write start control signal as well as FPGA select signals and several programming signals
to program the dynamic FPGA fast clock and the bit stream.
Whenever the host software selects an FPGA on a board, the micro-controller asserts the corresponding select line in order to
conduct subsequent bus communication or programming actions.

\section{Evaluation}
\label{sec:WPA2ClusterWPA2ClusterEvaluation}

We performed multiple evaluations with regard to our design performance, the power usage
and performance in comparison to GPUs as well as a Wi-Fi network security evaluation in
the form of a case study.

\subsection{FPGA Performance and Power Evaluation}
\label{subsec:WPA2ClusterPerformanceEvaluation}

\begin{figure}
    \captionsetup[subfigure]{labelformat=empty}
    \centering
    \subfloat{{\includegraphics[width=0.39\linewidth]{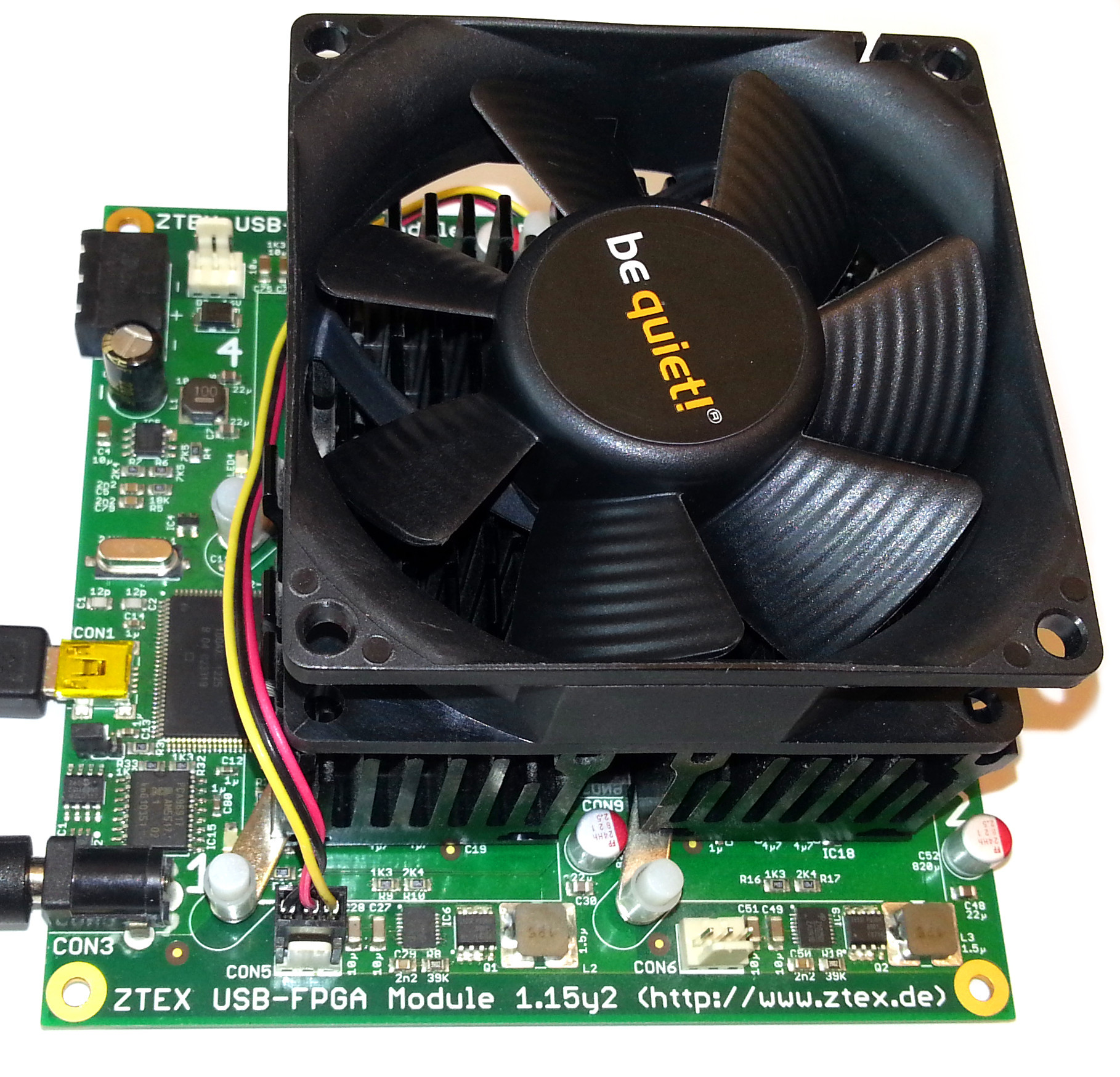} }}
    \qquad
    \subfloat{{\includegraphics[width=0.43\linewidth]{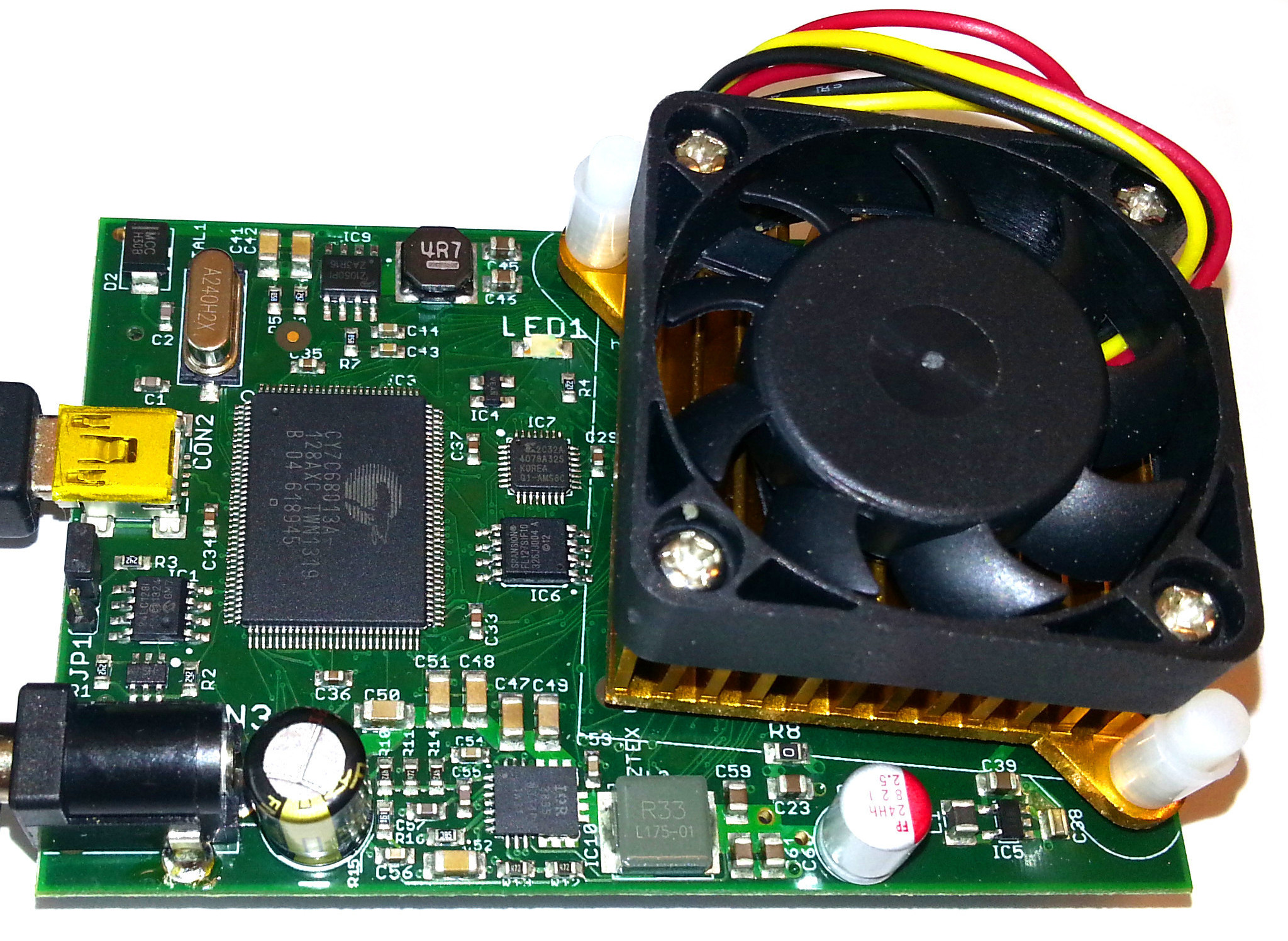} }}
    \caption{Ztex 1.15y Board (left), Ztex 2.16 Board (right)}
    \label{fig:WPA2ClusterFPGABoards}
\end{figure}

We evaluated the performance and the power usage of our design on multiple FPGAs and FPGA boards.
The first FPGA we targeted was an older model Spartan-6 XC6SLX150T-3 device. Four of these FPGAs
can be found on the Ztex 1.15y board visible on the left side of Figure \ref{fig:WPA2ClusterFPGABoards}.
The second FPGA we used for our evaluation was an Artix-7 XC7A200T-2 device on the Ztex 2.16 board
visible on the right side of the picture.
For both FPGAs, we created an optimized implementation and a configuration bit stream that can be uploaded to the device.
The main difference between the bit streams is the FPGA type, the maximum clock frequency and most importantly the
number of brute force cores we were able to fit onto the device.

\begin{figure}[htbp]
\centering
\includegraphics[width=0.8\textwidth]{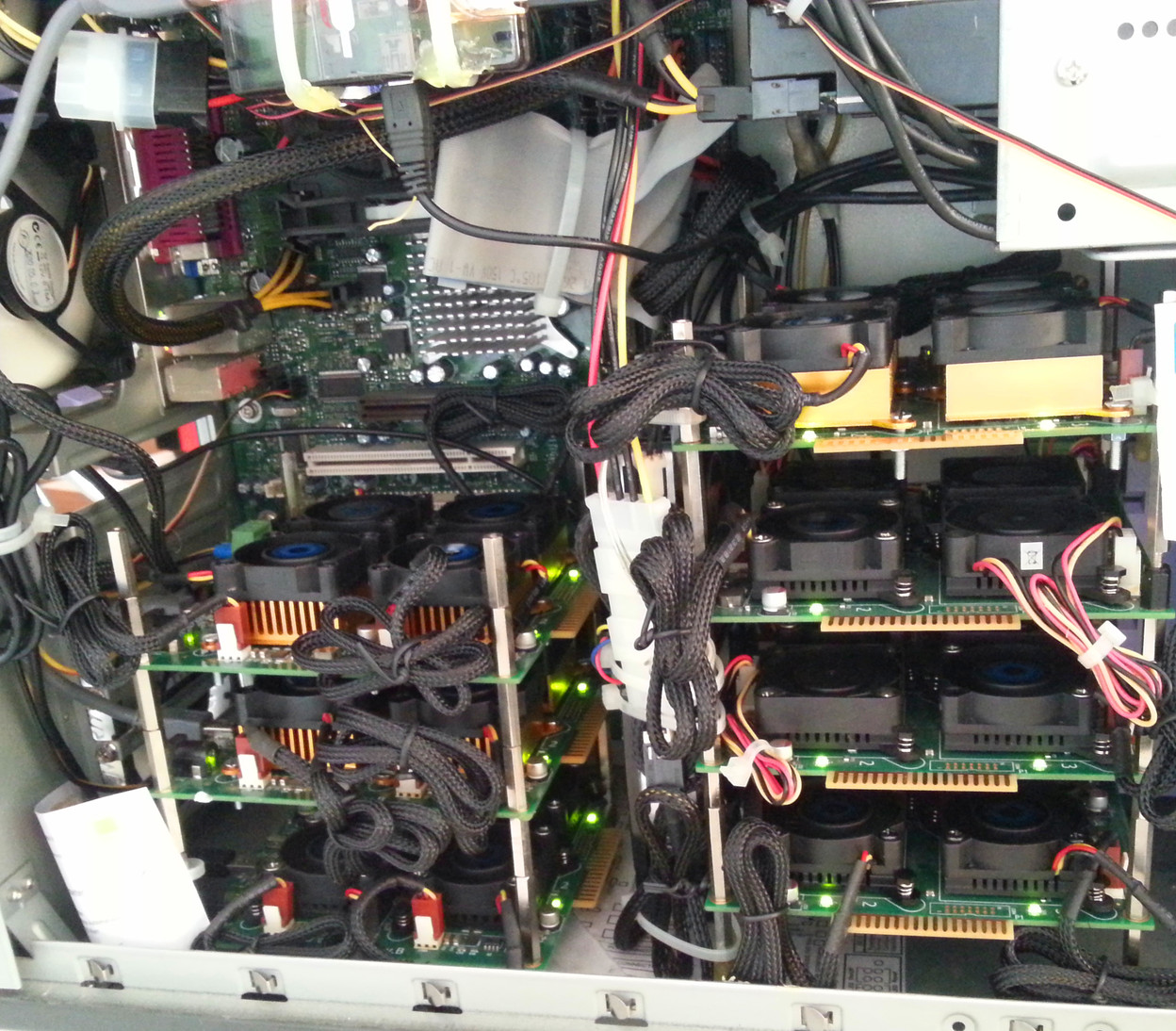}
\caption{Spartan-6 XC6SLX150T Cluster}
\label{fig:WPA2ClusterSpartan6Cluster}
\end{figure}

To evaluate the performance and the power requirements, we used the obtained timing and power
reports by utilizing the Xilinx timing and power analysis tools.
In addition to these results, we also conducted practical measurements on the FPGA boards.
At first, we measured the idle wattage of each unconfigured board at the power supply
to determine the idle power usage. In the next step, we used a generated WPA2-Personal handshake
with our software to mount a brute force attack on each of our FPGA boards.
We used large password working packages resulting in a $30$ seconds runtime per FPGA to avoid I/O bottlenecks.
By measuring the wattage again during operation, we were able to determine the overall power consumption.
To reduce the influence of the power consumption caused by losses in the power supplies or components other than the FPGA,
we obtained the power consumption of our FPGA implementation through the difference
between the overall idle consumption and the consumption during operation.
In Section \ref{subsec:WPA2ClusterGPUComparison}, we use the same method to determine the power consumption of GPUs to
get results that can be compared to the FPGA power consumption.
To obtain brute force performance measurements as well, we let each system run for at least 1 hour and
computed the performance by measuring the number of password guesses during that time. The result
is the average number of password guesses per second.
In addition to these evaluations, we executed the implementation on our FPGA cluster with $36$ Spartan-6 XC6SLX150T FPGAs
located on $9$ Ztex 1.15y FPGA boards. The cluster setup allowed us to perform measurements on a larger setup and to
determine how well our design scales with an increasing number of FPGAs.
The setup is visible in Fig. \ref{fig:WPA2ClusterSpartan6Cluster} whereat
two of those boards are not inside of the cluster as we use them for development purposes. During the tests, they were connected
externally to the cluster.
Using the power and performance measuring methodology from above, we obtained measurement results for the cluster as well.
To allow comparison with the commercial Elcomsoft WPA2-Personal FPGA cluster password recovery system \cite{Elcomsoft, PicoComputing},
we created an implementation and a configuration bitstream for the more expensive Kintex-7 XC7K410T-3 devices as well.
However, since we did not have a board with this type of Kintex-7 FPGA, we can provide the Xilinx development tool's timing and power analysis
results only.

\subsection{GPU Comparison}
\label{subsec:WPA2ClusterGPUComparison}

To measure performance and power requirements of GPUs, we utilized cudaHashcat\footnote{\url{http://hashcat.net/oclhashcat}} v1.36
to mount brute force attacks on the same WPA2-Personal handshake we used previously to test our FPGA implementations.
We executed the tool on machines with different Nvidia GPUs (GeForce GTX 750 Ti, GeForce GTX770 Windforce OC, GRID K520) and measured the performance in passwords per second
as well as the power consumption. We applied the same power measurement methodology as during our FPGA evaluation.
For the Amazon EC2 GPU cloud machines with GRID K520 GPUs, we were unable to obtain power measurements. The specific
machine configurations and results are described in detail in Section \ref{sec:WPA2ClusterResults}.

\subsection{Wi-Fi Security Evaluation - A Case Study}
\label{subsec:WPA2ClusterWiFiSecurityEvaluation}

\begin{figure}[htbp]
\centering
\includegraphics[width=0.7\textwidth]{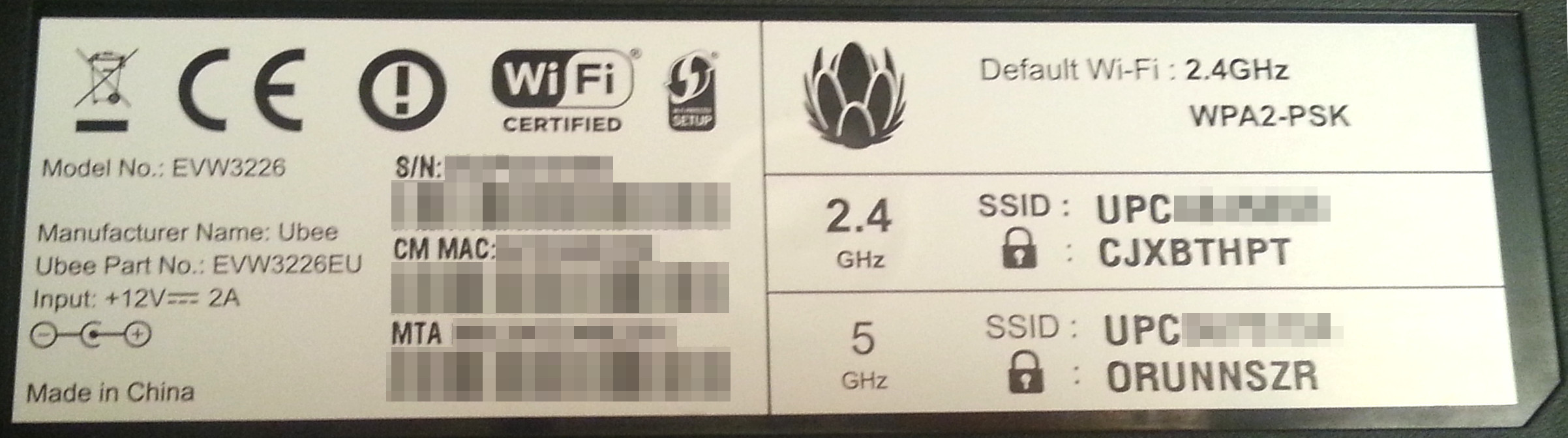}
\caption{Bottom Side of a Cable Modem}
\label{fig:WPA2ClusterCableModem}
\end{figure}

Driven by the high brute force speeds that can be achieved with FPGAs, we wanted
to evaluate whether there is a real-world security impact. While long random passwords
with a significantly large character set are practically infeasible to break within
a reasonable time frame, the minimum WPA2-Personal password length is only 8 characters \cite{citeulike:12556290}.
If the character set is limited as well, a random password can fall victim to brute force attacks
within days or even hours if the brute force speed is high enough.
To our surprise, discussions within our group suggested that the default WPA2-Personal passwords for
many mobile Wi-Fi modems and even ISP provided modems/routers not only have a limited character set
such as uppercase letters only, but the length of the variable part of the password or the length of the password
itself is also not more than 8 characters.
Further investigation turned out that the largest ISP in our country uses weak default passwords for many of its Wi-Fi cable modems
with only 8 characters length and comprising only uppercase letters. An example is provided in Fig. \ref{fig:WPA2ClusterCableModem}.
The manual of the Wi-Fi enabled cable modem further confirmed our finding \cite{UPCModem}.

While users can change these settings, the default \texttt{SSID} is \texttt{UPC<n>} where \texttt{n} denotes
a number with either 6 or 7 digits. Under the assumption that most users also change the \texttt{SSID} of
their network to something easier to remember when changing their Wi-Fi password, any visible \texttt{UPC<n>} Wi-Fi network
would be an indication that the network is still using the weak default password.

To evaluate the practical security impact of our implementation, we used the Wigle war driving dataset \cite{Wigle}
to get an approximation of how many of those likely to be insecure networks exist and whether those networks are limited to our country.
To do so, we created a rectangular scanning grid across the country and queried the Wigle Web service \cite{Wigle}. Since the
rectangular scanning grid also included parts of neighboring countries as well, it allowed us to see whether
the same ISP is active in those countries and the potentially weak cable modems are used there as well.
The results of our case study are presented in Section \ref{subsec:WPA2ClusterCaseStudyResults}.

\section{Results and Discussion}
\label{sec:WPA2ClusterResults}

In the following, we present the results of the performance and power evaluation of our FPGA implementations,
we present the obtained GPU WPA2-Personal brute force performance and power measurements results and comparison as well as
the outcome of the Wi-Fi security evaluation case study.

\subsection{FPGA Performance and Power Results}
\label{subsec:WPA2ClusterFPGAResults}

\begin{table*}[!htbp]
\begin{center}

\scalebox{0.68}{
  \begin{tabular}{ | l| r | r | r | r | r | r | r | r | r | r | r |}
  \hline
  System          &  FPGAs  &   Type       &   Cost  & Cores & Tool W    & Tool MHz & Meas. W & Act. MHz & calc pwd/s  &   pwd/s  & pwd/s W \\
  \hline
  Ztex 1.15y      &   1     & XC6SLX150T-3 &     175 &  2    &    4.281  & 187      &   6.99* & 180      &     21,956  &  21,871  & 3,128* \\
  Ztex 1.15y      &   4     & XC6SLX150T-3 &     700 &  8    &   17.124  & 187      &  27.96  & 180      &     87,826  &  87,461  & 3,128 \\	
  9x Ztex 1.15y   &  36     & XC6SLX150T-3 &   2,400 & 72    &  154.116  & 187      & 254     & 180      &    790,436  & 741,200  & 2,918 \\	
  Ztex 2.16       &   1     & XC7A200T-2   &     213 &  8    &   10.458  & 180      &  11.04  & 180      &     87,826  &  87,737  & 7,947 \\    
  N/A             &   1     & XC7K410T-3   &   2,248 & 16    &   25.634  & 216      &   N/A   & N/A      &    210,783  &     N/A  & N/A   \\
  N/A             &  48     & XC7K410T-3   & 107,904 & 768   & 1,230.432 & 216      &   N/A   & N/A      & 10,117,584  &     N/A  & N/A   \\
  \hline
  \end{tabular}
}
\caption{Performance and Power Results of our Implementations for different FPGA Devices and Systems/Boards}
\label{tab:WPA2ClusterFPGAResults}
\end{center}
\end{table*}

The results for our FPGA performance and power evaluation are visible in Table \ref{tab:WPA2ClusterFPGAResults}.
In the \texttt{System} and \texttt{FPGA} column the table shows on which systems we conducted our tests and how many FPGAs there are
on the corresponding board and/or in the overall system. The FPGA device types are visible in the \texttt{Type} column whereat
the name before the hyphen is the Xilinx device name and the number after the hyphen indicates the device speed grade (the higher the better).
The \texttt{Cost} column provides an approximate cost estimate per FPGA in US\$ we obtained by looking up the devices at common Xilinx distributors such as Digi-key\footnote{\url{http://www.digikey.com}}.
However while the cost for $9$ new Ztex 1.15y would be appoximately $6,300$ US\$, we considered our $9$ second-hand Ztex 1.15y boards previously used for cryptocurrency mining instead.
We were able to obtain these boards for $2,400$ US\$ which we believe is what amateurs could do as well, depending on how much boards they would like to acquire and how much they are willing to spend.
The \texttt{Cores} column shows how many cores we were able to fit onto the device to achieve maximum performance. While more cores per device generally increase
the performance, it can also cause the maximum clocking speed to drop significantly due to mapping, placement and routing issues. The table presents the implementations
allowing us to achieve the maximum performance per device.
The \texttt{Tool W} and \texttt{Tool MHz} columns present the design tool's power and timing analysis results. For the Spartan-6 FPGAs, we used the Xilinx ISE Suite 14.7 whereas
for the newer 7-series devices Artix-7 and Kintex-7, we used Vivado Design Suite 2015.1. In general, it appeared that the newer Vivado tools produced better results, but since it doesn't support
older model 6-series devices, we were unable to use it for our Spartan-6 implementations.
The \texttt{Meas.~W} and \texttt{Act.~MHz} columns present the results for the power measurements we conducted on the FPGA boards/systems and the actual clock speed we used to run the devices.
The \texttt{calc pwd/s} and \texttt{pwd/s} columns provide the WPA2-Personal performance in passwords per second whereas the first one indicates the calculated
and theoretic maximum performance of our implementation whereas the latter one shows the actual measured average performance per board and/or system.
In the last column \texttt{pwd/s W}, we use our actual power and performance measurements to determine how much brute force speed can be achieved per Watt which is
especially important when scaling up our implementation to larger FPGA cluster systems.
In the following, we discuss the results of our implementations on a per-device basis.

\subsubsection{Spartan-6 Results}
\label{subsubsec:WPA2ClusterSpartan6Results}

We used the Xilinx Spartan-6 XC6SLX150T-3 FPGA as the target for our initial implementation
due to the availability of a high-performance FPGA cluster with 36 of these devices at our lab.
The implementation on the Spartan-6 turned out to be especially challenging
for multiple reasons. We had to deal with long design tool runs (3 hours of more) each
time we made modifications to the design. Since the effects of many of our optimizations could not be tested through
behavioral simulations alone, the duration of the design tool runs significantly slowed down
the development. In addition, the internal switch boxes and types of slices in the Spartan-6 architecture
are not well suited for more complex and larger implementations in comparison to newer 7-series devices.
The result was that for many of our implementation attempts the device logic resources were sufficient, but
the implementation still turned out to be unroutable due to the number of required interconnects. An important factor
to achieve routable designs was our use of FPGA floor planning.

\begin{figure}[htbp]
\centering
\includegraphics[width=0.6\textwidth, angle=90]{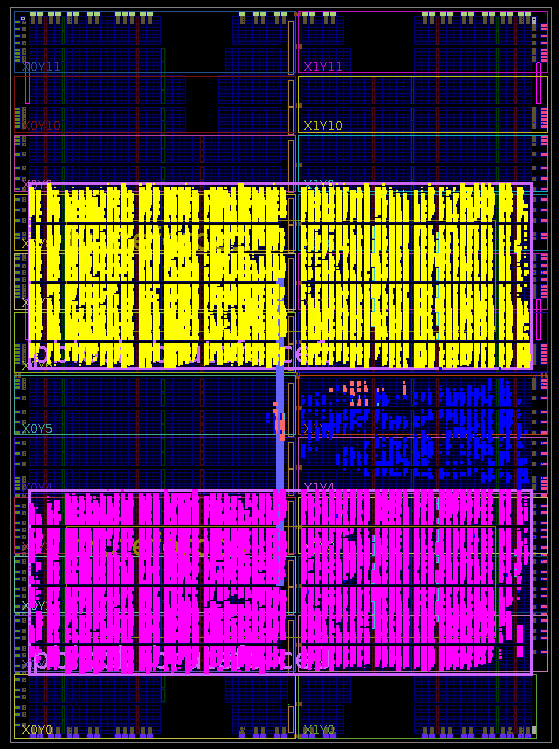}
\caption{Placed and Routed XC6SLX150T}
\label{fig:WPA2ClusterSpartan6Implementation}
\end{figure}

In summary, we were able to generate two implementations for the Spartan-6 XC6SLX150T-3. One with 3 cores and one with 2 cores.
While the first one has an additional core in comparison, it resulted in a much lower achievable clock speed (62.5 MHz)
due to placing and routing issues effectively reducing the performance to that of a single core at high speed (180 MHz).
In contrast, our optimized 2 core variant visible in Fig. \ref{fig:WPA2ClusterSpartan6Implementation} is able to
run at up to $187$ MHz leading to the highest performance we were able to achieve on the device.
The picture shows the ready-to-upload placed and routed design. On the left and right the 2 brute force cores are clearly visible.
In between the password generator and the global state machine are located. Although the dark areas indicate that
there would be sufficient space for an additional core, our experiments showed that this would lead to lower performance
as explained above.

The first 3 rows in Table \ref{tab:WPA2ClusterFPGAResults} present the results we obtained through this implementation.
Due to cooling requirements, we ran the design with a reduced clock speed of $180$ MHz.
Our measurements indicate that in this configuration, our implementation requires a total of $27.96$W for all $4$ FPGAs
on the Ztex 1.15y board. The power measurements per Spartan-6 FPGA are marked with an asterisk to indicate that 
we were unable to measure them directly, but rather derived the measurement results from our power measurements for the
entire Ztex 1.15y board with its $4$ FPGAs.
Our results show that our approach scales well and can be easily run in a cluster configuration producing
a performance of $790,436$ password guesses per second on our cluster.
The difference between the calculated maximum performance and the measured performance is mainly due
to the I/O times between the PC, the microcontroller and the FPGAs.
In addition, our Spartan-6 implementation includes a dynamic frequency scaling mechanism slowing down the
FPGAs in case of device temperatures getting too high. With better cooling inside the cluster, we believe that
the gap between the theoretic performance and the measured performance could be made smaller.

\subsubsection{Artix-7 Results}
\label{subsubsec:WPA2ClusterArtix7Results}

\begin{figure}[htbp]
\centering
\includegraphics[width=0.6\textwidth, angle=90]{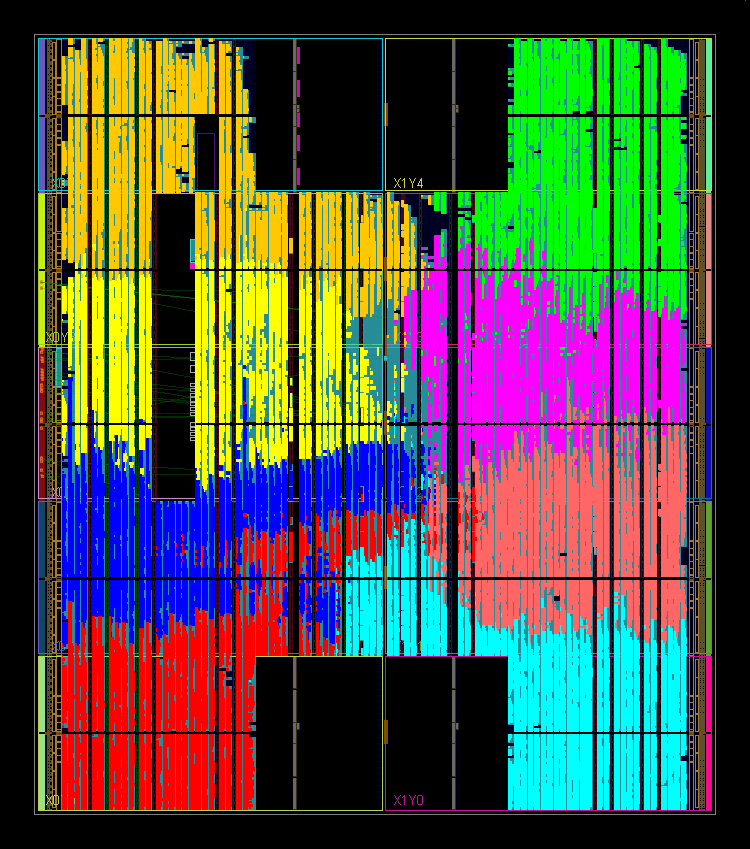}
\caption{Placed and Routed XC7A200T}
\label{fig:WPA2ClusterArtix7Implementation}
\end{figure}

In comparison to our Spartan-6 implementations, our implementations on the newer 7-series Artix-7 XC7A200T-2 FPGA
required less effort as we could not only start from our already highly optimized Spartan-6 design, but the architecture and the
newer Vivado design tool are also better suited for larger designs with increasing design complexity.
Since device internals such as the clocks or PLLs are different from the Spartan-6 architecture, we had to
adapt our implementation accordingly. The ability to read the device's core temperature from within the FPGA implementation
was especially interesting. It allowed us to implement frequency scaling mechanisms directly on the FPGA not only preventing possible damage
due to overheating, but also ensuring that each device always runs at the maximum performance possible.
While we don't have access to an Artix-7 FPGA cluster, this feature would be especially helpful for high-performance cluster designs.

Our ready-to-upload placed and routed design is visible in Fig. \ref{fig:WPA2ClusterArtix7Implementation}. The black blocks on the left and right
are unusable areas. As routing around those areas makes it hard to meet timing constraints, we utilized floor planning
to provide approximate locations for all of the $8$ cores we managed to fit onto the device.
All of the cores have a small path to the center where the small block with the global state machine and the password generator are located.
The implementation can be run at up to $180$ MHz to achieve a theoretic maximum of $87,826$ password guesses per second.
We managed to create an implementation with $9$ cores as well, but similar to our Spartan-6 implementations the overall performance
would have dropped due to the lower maximum clock frequency caused by placing and routing issues.
With a measured performance of $87,737$ password guesses per second, our results show that a single XC7A200T-2 device achieves not only
more performance than 4 of the older model Spartan-6 XC6SLX150T-3 FPGAs altogether, but it also requires just $11.04$ Watt during operation.

\subsubsection{Kintex-7 Results}
\label{subsubsec:WPA2ClusterKintex7Results}

\begin{figure}[htbp]
\centering
\includegraphics[width=0.6\textwidth, angle=90]{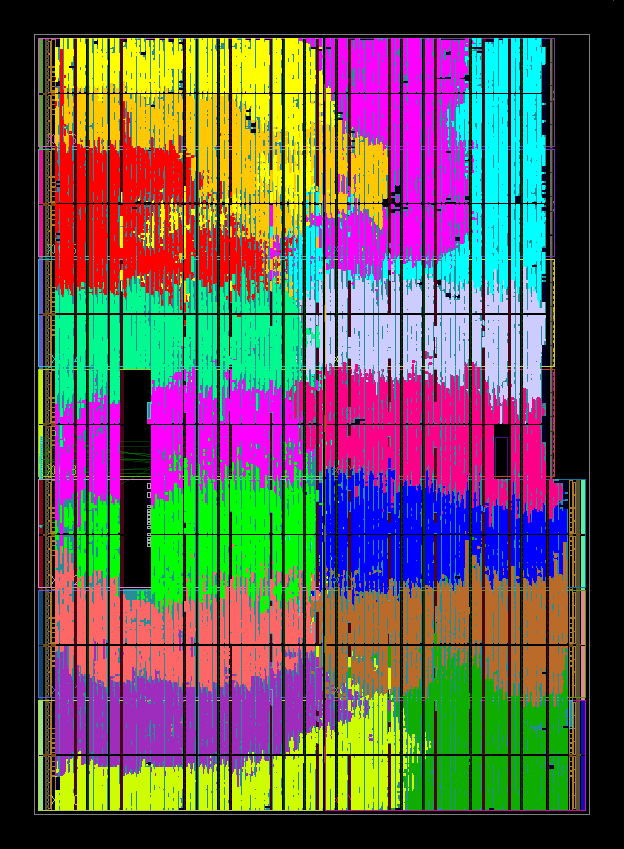}
\caption{Placed and Routed XC7K410T}
\label{fig:WPA2ClusterKintex7Implementation}
\end{figure}

In contrast to the low-cost Artix-7 FPGAs, Kintex-7 FPGAs are larger and allow higher performance but are also significantly more expensive.
Although we didn't have any of those FPGAs at hand, we created an implementation for the Kintex-7 XC7K410T FPGA for two reasons.
First, Elcomsoft's marketed to be world's fastest FPGA-based WPA2 password recovery system relies on these FPGAs just the same
and even provides performance figures for it \cite{PicoComputing}. Our targeting of the same FPGAs thus allows direct performance
comparison between their implementation and ours. Their document indicates that on the PicoComputing SC5/M505-48 cluster
with 48 XC7K410T FPGAs their implementation is able to produce $1,988,360$ passwords guesses per second \cite{PicoComputing}.
Assuming that their implementation targets WPA2 employing SHA1 instead of WPA1 employing the much less complex MD5 algorithm,
our implementation could achieve up to $10,117,584$ passwords per second on the same hardware and would thus be more than 5 times as fast.
Second, we wanted to obtain performance data for larger FPGAs as well.
Although expensive, we believe that Kintex-7 FPGAs are well in the price range for professional attackers
allowing them to achieve significantly more brute force attack performance per FPGA in comparison to low-cost FPGAs such as the Artix-7.
Our ready-to-upload placed and routed design is visible in Fig. \ref{fig:WPA2ClusterKintex7Implementation}. It comprises $16$ cores
running at up to $216$ MHz. Similar to our Artix-7 implementation, the password generator and the global state machine are located in the center.
However, due to the size and the thin layout of those units, they are hardly visible in the picture.
At the same time, the image also suggests that with an increasing number of cores, the centralized state machine
and password generator becomes a bottleneck due to the long bus interconnects reaching to the outside cores.
We believe that this problem could be easily addressed by either using multiple shared state machines and password generators or
by including FIFOs for the password candidates in each of the brute force cores.
Due to the long runtime of each brute force core, the FIFO could be filled with a slow clock that
can be easily routed across long distances on the FPGA. At the same time, the brute force cores
would operate on the fast clock and drain the FIFOs. Due to their long run time, the FIFOs could be easily re-filled
through the slow clock before the next set of password candidates would be required.

\subsection{GPU Results and Comparison}
\label{subsec:WPA2ClusterGPUResults}

\begin{figure*}[h!tbp]
    \captionsetup[subfigure]{labelformat=empty}
    \centering
    \subfloat{{\includegraphics[width=0.46\linewidth]{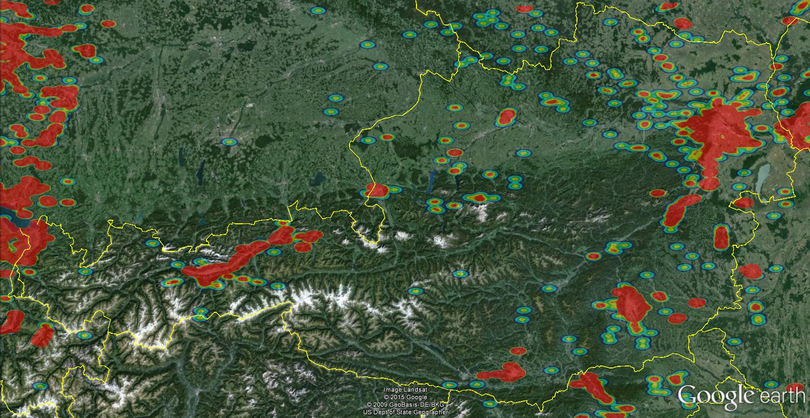} }}
    \qquad
    \subfloat{{\includegraphics[width=0.46\linewidth]{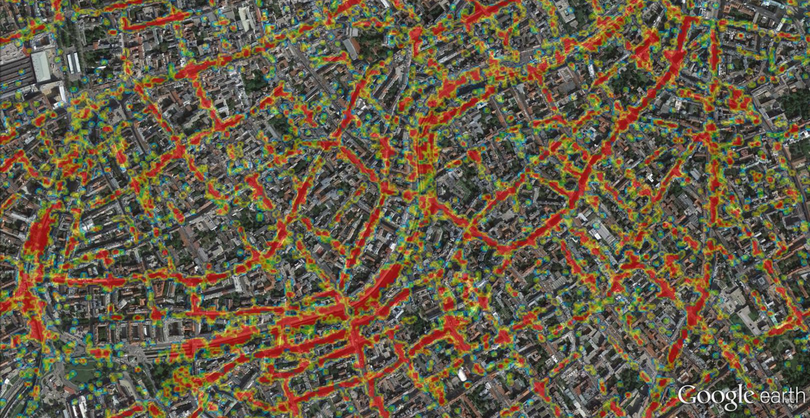} }}
    \caption{Density of \texttt{UPC<n>} networks with potentially weak WPA2-Personal Passwords}
    \label{fig:WPA2ClusterMapDensity}
\end{figure*}

\begin{table}[!htbp]
\begin{center}

\scalebox{1.0}{
  \begin{tabular}{ | l| r | r | r |}
  \hline
  System                    &  pwd/s    &   W           &  pwd/s W\\
  \hline
  GeForce GTX750 Ti         & 52,446    & 106           & 495  \\
  GeForce GTX770 OC         & 62,420    & 184           & 339  \\
  Amazon EC2 - GRID K520    & 30,370    & N/A           & N/A  \\
  Amazon EC2 - GRID K520 x4 & 109,073   & N/A           & N/A \\ 
  \hline
  \end{tabular}
}
\caption{Performance and Power Results on GPUs}
\label{tab:WPA2ClusterGPUResults}
\end{center}
\end{table}

The results of our GPU evaluation (Section \ref{subsec:WPA2ClusterGPUComparison}) are visible in Table \ref{tab:WPA2ClusterGPUResults}.
We performed the performance measurements by running cudaHashcat v1.36 on different systems and measuring
the power consumption as the difference between idle and busy WPA2 computations
to get results independent from other components in the system.
The table shows the different GPU configurations (\texttt{System}) we used for our tests.
The \texttt{pwd/s} column shows the performance in passwords per second and the \texttt{W} column indicates
the power consumed by the GPU during runtime in Watt. The performance per Watt is visible in the \texttt{pwd/s W} column.

In addition to running GPU measurements on our own machines, we also conducted measurements on dedicated Amazon Elastic Cloud (EC2) GPU machines as well.
While we could measure the performance on the machines just the same, we were unable to obtain power measurements.
Although using a high number of GPU cloud machines appears promising to achieve high brute force attack performance, the limiting factor
is the cost. Although our combined experiments on the dedicated Amazon EC2 machines took no longer than an hour, the costs we
accumulated for our tests were already US\$ $14.92$.
Since realistic brute force attacks might take considerably longer, the costs for an attacker would be
far lower for acquiring a powerful GPU system instead of using the Amazon EC2 GPU nodes.
In comparison to the results we obtained from our FPGA implementation, it is visible that GPUs
can achieve the performance of a state-of-the-art low-cost FPGA (i.e., Artix-7), but their power consumption
and performance per Watt is more than $10$ times as high. At the same time, the performance achievable with a single larger
FPGA such as the Kintex-7 XC7K410T is no longer in the range of GPUs.
Considering high-speed attacks with clusters, we believe that the scalability for FPGA-based attacks is better as well
due to the small size of FPGAs, their lower power consumption and the high performance they can produce.

\subsection{Wi-Fi Security Case Study}
\label{subsec:WPA2ClusterCaseStudyResults}

\captionsetup[table]{skip=5pt}
\begin{table}[htbp]
\begin{center}

\scalebox{1.0}{
  \begin{tabular}{ | l| r | }
  \hline
  Area                     &  Networks \\
  \hline
  Vienna                   & 120,380	\\ 
  Austria + Border region  & 166,988\\
  \hline
  \end{tabular}
}
\caption{UPC Networks with likely weak WPA2-Personal Passwords}
\label{tab:WPA2ClusterUPCResults}
\end{center}
\end{table}

The results for our real-world security evaluation case study are visible in Table \ref{tab:WPA2ClusterUPCResults}.
Within the city of Vienna,
we found $120,380$ Wi-Fi networks with the \texttt{SSID} having the form \texttt{UPC<n>} in the Wigle \cite{Wigle} dataset.
We were astonished by the high density of these networks within the city (Fig. \ref{fig:WPA2ClusterMapDensity} on the right), but assume that the real number
of networks is even higher as not all networks are covered in the Wigle dataset.
In addition, we discovered that the security issue is not only limited to the city of Vienna,
but it is also persistent in the whole country
and even in the border regions to neighboring countries. The left picture of Fig. \ref{fig:WPA2ClusterMapDensity} provides an overview of the
network density. In total, we discovered $166,988$ of these networks within that area including neighboring countries.
Due to weak default configuration (Section \ref{subsec:WPA2ClusterWiFiSecurityEvaluation}), our case study suggests
that our FPGA cluster implementation could be used to break into each of those networks in no more than $3$ days considering
the rate of $790,436$ guesses per second and the small number of only $26^8$ password combinations for each vulnerable network.
Running our implementation on the PicoComputing SC5/M505-48 cluster \cite{PicoComputing} instead,
the necessary worst-case time to break a network would be further reduced down to $5.7$ hours.
Due to the severe security implications and the high number of private networks involved,
we already reported the problem to the ISP and are currently in the process of reporting it to the national CERT team as well.

\section{Conclusion and Future Work}
\label{sec:WPA2ClusterConclusionAndFutureWork}

In this paper, we demonstrated that WPA2 passwords can be attacked at high speed rates
not only by expensive professional FPGA cluster solutions but similar speeds can be achieved by amateurs
on a budget as well, especially when considering second hand FPGA boards previously used for cryptocurrently mining.
We specifically targeted low-cost FPGA devices, conducted implementations on 3 different FPGA architectures
and evaluated our results with regard to performance and power.
Our GPU evaluation suggests that FPGAs can not only achieve higher speeds
at significantly less power, but they can also be used to easily create small and afforable FPGA clusters
in the reach of amateurs.
We conducted a real-world security evaluation showing that within the country and its border regions, there
are more than $166,000$ Wi-Fi networks with likely weak WPA2-Personal passwords
that could be attacked through the implementation on our FPGA cluster within no more than 3 days each.
Considering commercially used FPGA cluster systems, the time could be further reduced
to no more than $5.7$ hours depending on the cluster configuration and device types.
However, we believe that besides the speedup we achieved it is more important to consider 
that the WPA2-Personal brute force performance achievable on professional systems is now
becoming feasible on the low-cost systems amateurs can afford as well.
We believe that these low-cost FPGA cluster based brute force attacks are
thus a serious threat to real-world systems and need to be especially considered by manufacturers
when choosing WPA2-Personal default passwords for hundred thousands of devices.
As counter measure, users need to increase the length of their passwords,
the password should be random and it should utilize a large character set to increase password entropy.
In future work, we are looking forward to evaluate the security of other
cryptographic systems as well. In that regard, we plan to
design and implement a powerful low-cost FPGA cluster similar to COPACOBANA \cite{HiPerfCOPACOBANA}
but with low-cost 7-series devices instead.

\section*{Acknowledgments}
The research was funded by the Austrian
Research Funding Agency's (FFG) KIRAS security research
program through the (SG)$^2$ project under national FFG grant
number 836276, the AnyPLACE project under EU H2020
grant number 646580, and the IT security consulting company
Trustworks KG who also provided the FPGA boards and the cluster.


\balance
\bibliographystyle{splncs03}
\bibliography{paper}

\begin{thebibliography}{10}
\providecommand{\url}[1]{\texttt{#1}}
\providecommand{\urlprefix}{URL }

\bibitem{Bittau:2006:FNW:1130235.1130389}
Bittau, A., Handley, M., Lackey, J.: The final nail in wep's coffin. In:
  Proceedings of the 2006 IEEE Symposium on Security and Privacy. pp. 386--400.
  SP '06, IEEE Computer Society, Washington, DC, USA (2006)

\bibitem{rfc3174}
{{D. Eastlake 3rd and P. Jones}}: {{US Secure Hash Algorithm 1 (SHA1)}}. RFC
  3174 (Informational) (Sep 2001), updated by RFCs 4634, 6234

\bibitem{rfc6234}
{{D. Eastlake 3rd and T. Hansen}}: {US Secure Hash Algorithms (SHA and
  SHA-based HMAC and HKDF)}. RFC 6234 (Informational) (May 2011)

\bibitem{ElcomsoftFastest}
{Elcomsoft}: {ElcomSoft and Pico Computing Demonstrate World's Fastest Password
  Cracking Solution}. \url{https://www.elcomsoft.com/PR/Pico_120717_en.pdf},
  [Online; accessed 13-Nov-2015]

\bibitem{Elcomsoft}
{Elcomsoft Blog}: {Accelerating Password Recovery: the Addition of FPGA}.
  \url{http://blog.elcomsoft.com/2012/07/accelerating-password-recovery-the-addition-of-fpga}
  (2012), [Online; accessed 13-Nov-2015]

\bibitem{HiPerfCOPACOBANA}
Güneysu, T., Kasper, T., Novotný, M., Paar, C., Wienbrandt, L., Zimmermann,
  R.: {High-Performance Cryptanalysis on RIVYERA and COPACOBANA Computing
  Systems}. In: Vanderbauwhede, W., Benkrid, K. (eds.) {High-Performance
  Computing Using FPGAs}, pp. 335--366. Springer New York (2013),
  \url{http://dx.doi.org/10.1007/978-1-4614-1791-0_11}

\bibitem{citeulike:12556290}
{IEEE-Inst.}: {802.11-2012 - IEEE Standard for Information
  technology--Telecommunications and information exchange between systems Local
  and metropolitan area networks--Specific requirements Part 11: Wireless LAN
  Medium Access Control (MAC) and Physical Layer (PHY) Specifications}. Tech.
  Rep. IEEE Std 802.11™-2012, IEEE-Inst (2012),
  \url{http://ieeexplore.ieee.org/servlet/opac?punumber=6178209}

\bibitem{JohRog15A}
Johnson, T., Roggow, D., Jones, P., Zambreno, J.: An fpga architecture for the
  recovery of wpa/wpa2 keys. Journal of Circuits, Systems, and Computers (JCSC)
   24(7) (2015)

\bibitem{rfc2898}
Kaliski, B.: {PKCS \#5: Password-Based Cryptography Specification Version 2.0}.
  RFC 2898 (Informational) (Sep 2000),
  \url{http://www.ietf.org/rfc/rfc2898.txt}

\bibitem{5234856}
Lashkari, A., Danesh, M., Samadi, B.: A survey on wireless security protocols
  (wep, wpa and wpa2/802.11i). In: Computer Science and Information Technology,
  2009. ICCSIT 2009. 2nd IEEE International Conference on. pp. 48--52 (Aug
  2009)

\bibitem{citeulike:1855432}
Oechslin, P.: {Making a Faster Cryptanalytic Time-Memory Trade-Off}. In: Boneh,
  D. (ed.) Advances in Cryptology - CRYPTO 2003, Lecture Notes in Computer
  Science, vol. 2729, chap.~36, pp. 617--630. Springer Berlin / Heidelberg,
  Berlin, Heidelberg (2003),
  \url{http://dx.doi.org/10.1007/978-3-540-45146-4\_36}

\bibitem{PicoComputing}
{PicoComputing Inc.}: {SC5-4U Overview}.
  \url{http://picocomputing.com/brochures/SC5-4U.pdf}, [Online; accessed
  13-Nov-2015]

\bibitem{JMEDSWPA_WPA2_Password_Security_Testing_using_Graphics_Processing_Units}
{Sorin Visan}: {WPA/WPA2 Password Security Testing using Graphics Processing
  Units}. Journal of Mobile, Embedded and Distributed Systems  5(4) (2013)

\bibitem{BreakingWEP}
Tews, E., Weinmann, R.P., Pyshkin, A.: Breaking 104 bit wep in less than 60
  seconds. In: Kim, S., Yung, M., Lee, H.W. (eds.) Information Security
  Applications, Lecture Notes in Computer Science, vol. 4867, pp. 188--202.
  Springer Berlin Heidelberg (2007),
  \url{http://dx.doi.org/10.1007/978-3-540-77535-5_14}

\bibitem{UPCModem}
{Ubee}: {Ubee EVW3226 Advanced Wireless Voice Gateway - Bedienungsanleitung}.
  \url{http://www.unitymedia.de/content/dam/unitymedia-de/hilfe---service/doc/ubee-evw3226-installationsanleitung.pdf}
  (2011), [Online; accessed 22-April-2015]

\bibitem{NIST180-2}
{U.S.Department of Commerce National Institute of Standards and Technology}:
  {FIPS PUB 180-2, Secure Hash Standard (SHS)} (2002), {U.S.Department of
  Commerce/National Institute of Standards and Technology}

\bibitem{NIST140-2}
{U.S.Department of Commerce/National Institute of Standards and Technology}:
  {FIPS PUB 140-2, Security Requirements for Cryptographic Modules} (2002),
  {U.S.Department of Commerce/National Institute of Standards and Technology}

\bibitem{Vanhoef:2013:PVW:2484313.2484368}
Vanhoef, M., Piessens, F.: {Practical Verification of WPA-TKIP
  Vulnerabilities}. In: Proceedings of the 8th ACM SIGSAC Symposium on
  Information, Computer and Communications Security. pp. 427--436. ASIA CCS
  '13, ACM, New York, NY, USA (2013),
  \url{http://doi.acm.org/10.1145/2484313.2484368}

\bibitem{Wigle}
{Wigle.net Project}: \url{https://wigle.net}, [Online; accessed 13-Nov-2015]

\bibitem{Spartan6Family}
{Xilinx Inc.}: {DS160, v2.0, Spartan-6 Family Overview}.
  \url{http://www.xilinx.com/support/documentation/data_sheets/ds160.pdf}
  (2011), [Online; accessed 22-April-2015]

\bibitem{ZTEX}
{ZTEX Gmbh}: {Products}. \url{http://www.ztex.de}, [Online; accessed
  22-April-2015]

\end{thebibliography}
%



\end{document}